%% file: main.tex
\newif\ifsupp
\supptrue 

\documentclass[superscriptaddress,twocolumn,10pt,prl,floatfix,nofootinbib]{revtex4-1}

\input{glyphtounicode}
\usepackage[T1]{fontenc}
\usepackage[utf8]{inputenc}
\usepackage[english]{babel}
\usepackage{amsmath}  
\usepackage{amsfonts} 
\usepackage{graphicx} 
\usepackage[pdftex,bookmarks=true,bookmarksopen,bookmarksnumbered,
                colorlinks,
                linkcolor=blue,
                citecolor=blue]{hyperref}
\graphicspath{ {./} }
\usepackage{natbib}
\usepackage{array}
\usepackage[table]{xcolor}
\usepackage{xcolor}
\usepackage{braket}
\usepackage{booktabs}
\usepackage{multirow}
\usepackage{caption}
\usepackage{dsfont} 

\newcommand{\aref}[1]{\hyperref[#1]{Appendix~\ref*{#1}}}
\DeclareCaptionLabelSeparator{line}{\hspace{2pt}\bf{|}\hspace{2pt}}

\begin{document}

\captionsetup[table]{name={TABLE},labelsep=period,justification=centerlast,font=small}
\captionsetup[figure]{name={\bf{Fig.}},labelsep=line,justification=centerlast,font=small}
\renewcommand{\equationautorefname}{Eq.}
\renewcommand{\figureautorefname}{Fig.}
\renewcommand*{\sectionautorefname}{Sec.}


\title{On-demand electric control of spin qubits}

\author{Will Gilbert}
\altaffiliation[]{These co-authors had equal contributions to this work.}
\author{Tuomo Tanttu}
\altaffiliation[]{These co-authors had equal contributions to this work.}
\author{Wee Han Lim}
\author{MengKe Feng}
\author{Jonathan Y. Huang}
\author{Jesus D. Cifuentes}
\author{Santiago Serrano}
\author{Philip Y. Mai}
\author{Ross C. C. Leon}
\author{Christopher C. Escott}
\affiliation{School of Electrical Engineering and Telecommunications, The University of New South Wales, Sydney, NSW 2052, Australia}
\author{Kohei M. Itoh}
\affiliation{School of Fundamental Science and Technology, Keio University, Yokohama, Japan}
\author{Nikolay V. Abrosimov}
\affiliation{Leibniz-Institut für Kristallzüchtung, 12489 Berlin, Germany}
\author{Hans-Joachim Pohl}
\affiliation{VITCON Projectconsult GmbH, 07745 Jena, Germany}
\author{Michael L. W. Thewalt}
\affiliation{Department of Physics, Simon Fraser University, British Columbia V5A 1S6, Canada}
\author{Fay E. Hudson}
\author{Andrea Morello}
\author{Arne Laucht} 
\author{Chih Hwan Yang}
\author{Andre Saraiva}
\author{Andrew S. Dzurak}
\affiliation{School of Electrical Engineering and Telecommunications, The University of New South Wales, Sydney, NSW 2052, Australia}
\date{\today}

\begin{abstract}

\textbf{Once called a ``classically non-describable two-valuedness'' by Pauli \cite{pauli1925uber}, the electron spin is a natural resource for long-lived quantum information since it is mostly impervious to electric fluctuations and can be replicated in large qubit arrays in silicon, offering high-fidelity control \cite{yang2019silicon, xue2022quantum, noiri2022fast, madzik2022precision, mills2021twoqubit, zwerver2021qubits}. Paradoxically, one of the most convenient control strategies is the integration of nanoscale magnets to artificially enhance the coupling between spins and electric fields \cite{golovach2006electric, pioro2008electrically, yoneda2017quantumdot}, which in turn hampers the spin's noise immunity \cite{kha2015do} and adds architectural complexity \cite{boter2021spiderweb}. Here we demonstrate a technique that enables a \textbf{\textit{switchable}} interaction between spins and orbital motion of electrons in silicon quantum dots, without the presence of a micromagnet. The naturally weak effects of the relativistic spin-orbit interaction in silicon are enhanced by more than three orders of magnitude by controlling the energy quantisation of electrons in the nanostructure, enhancing the orbital motion. Fast electrical control is demonstrated in multiple devices and electronic configurations, highlighting the utility of the technique. Using the electrical drive we achieve a coherence time $\boldsymbol{T_{2,{\rm Hahn}}\approx50}$~\textmu s, fast single-qubit gates with $\boldsymbol{T_{\pi/2}=3}$~ns, and gate fidelities of 99.93\% probed by randomised benchmarking. The higher gate speeds and better compatibility with CMOS manufacturing, enabled by on-demand electric control, improve the prospects for realising scalable silicon quantum processors.}
\end{abstract}

\maketitle

The density of quantum dots in an array is set by the size of the electron wave functions and the consequent size and pitch of gate electrodes~\cite{gonzalez2020scaling}, but individualised high-fidelity control of electron spin qubits in silicon typically requires on-chip integration of much larger devices, such as micromagnets~\cite{pioro2008electrically, leon2020coherent} or stripline antennae~\cite{koppens2006driven}. Other spin qubit implementations, such as electrons in InAs nanowires~\cite{nadj2010spin} and holes in silicon~\cite{maurand2016cmos} and germanium~\cite{watzinger2018germanium,scappucci2020germanium,froning2021ultrafast}, have sufficient intrinsic spin-orbit coupling to enable localised, all-electrical control employing only the gate electrodes that are already used to define the quantum dots. However, the same spin-orbit coupling that enables direct electrical control also exposes the qubits to decoherence from electrical noise~\cite{kha2015do}. Furthermore, while some semiconductor fabrication plants have the capability to integrate non-silicon materials, electrostatic quantum dots using silicon CMOS technology offer the strongest prospect of leveraging the full potential for integration and miniaturisation of the most advanced transistor fabrication nodes~\cite{ieee2021irds}.

\begin{figure*}[ht!]
    \includegraphics[width = 1\textwidth, angle = 0]{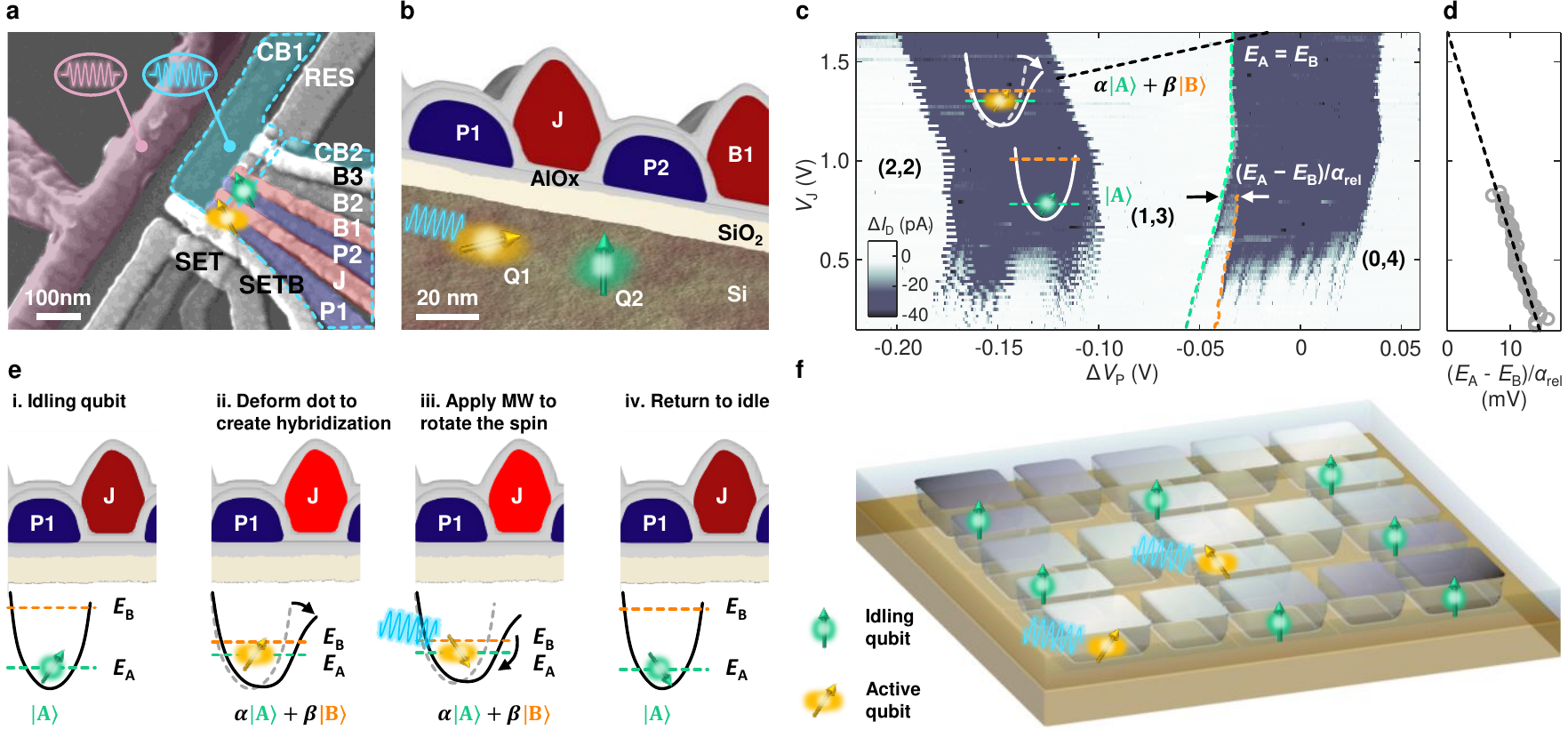}
    \caption{\textbf{Electrostatic quantum dots with tunable energy spectrum.} 
    \textbf{a}, Scanning electron micrograph of a quantum dot device nominally identical to all four devices investigated here. 
    \textbf{b}, Simulated cross section of the device geometry overlaid with a schematic of the electron wavefunctions and spins. 
    \textbf{c}, Excited state spectroscopy of an isolated double quantum dot containing 4 electrons, measured in device A as a function of $V_{\rm J}$ and the interdot voltage bias $\Delta V_{\rm P}$. Energies $E_{\rm A}$ (green dashed line) and $E_{\rm B}$ (orange dashed line) are fitted to step increases in interdot tunnel rate, indicating occurence of ground and first excited states.
    \textbf{d}, Extrapolation of the $E_{\rm A}-E_{\rm B}$ separation reveals a convergence near $V_{\rm J}=1.6$~V. 
    \textbf{e}, Control sequence illustrating how to deform the quantum dot to turn on a controllable orbital degeneracy using the lateral J gate and use it for qubit control. The idling state is purely spin-only, while the control state is a spin-orbit mix. 
    \textbf{f}, Rendering of a multi-qubit array with idling qubits set to be in a spin-only state, while some of the qubits are being controlled in a spin-orbit mixed state.}
    \label{fig:sem}
\end{figure*}

\subsection{Controlling the electron energy spectrum}
Spin-orbit effects for electrons in silicon quantum dots are typically small, meaning that direct electric driving is weak, however, these effects can become significant if the electron is allowed to move between orbital configurations within the quantum dot~\cite{leon2021bell}. These configurations are generally immutable due to the well quantised orbital energies of few-electron quantum dots. However, a dense arrangement of electrodes such as in \autoref{fig:sem}a and b (nominally identical to all devices studied here) gives access to a level of control over the potential landscape that can be used to consistently form quantum dots that possess an energy spectrum with flexible controllability. \ifsupp Further technical details on strategies to achieve this controllability are discussed in the Extended Data. \fi \autoref{fig:sem}c shows how excited state spectroscopy may be used to infer the presence of electronic states that have a differential lever arm $\alpha_{\rm rel}$, and correspond to different charge density distributions and which therefore couple differently to the various electrostatic gates. We denote these orbital configurations A and B, and their energies $E_{\rm A}$ and $E_{\rm B}$, respectively.

We instigate internal movement of the electron within the dot by biasing the gate voltages to reconfigure the quantum dot to a point where the two states have approximately the same energy. At this point, the quantum dot becomes highly polarisable, which leads to fast electrically-driven spin resonance (EDSR). This quasi-degeneracy point can be found by extracting the excitation energy (separation of fitted orange and green dotted lines in \autoref{fig:sem}c) extrapolating the trend against the side gate voltage (J gate) to the point where it reaches zero (\autoref{fig:sem}d). At that point, the A and B states hybridize, and the electron enters a superposition state $\alpha|{\rm A}\rangle+\beta|{\rm B}\rangle$. The exact values of $\alpha$ and $\beta$ depend on the particular nature of the two states A and B, but they are controllable by exploiting the differential lever arm $\alpha_{\rm rel}$.

This controllability over the wavefunction hybridization is the key to on-demand exploitation of spin-orbit effects. In \autoref{fig:sem}e we present a typical series of control steps, starting from an idling qubit (i) that is set to have minimal spin-orbit effects by setting the quantisation energy to be large ($\alpha\approx 1$ and $\beta\approx 0$). The dot is then deformed to create the hybrid state (ii) for a short amount of time, sufficient for the application of a microwave pulse (iii) that creates the spin rotation. The quantum dot is then reconfigured to the idling mode (iv), which restores the qubit resilience against spurious electric field fluctuations. This strategy allows for idling qubits to be protected for prolonged times while active qubits are being manipulated. The vision of a scalable qubit arrangement presented in \autoref{fig:sem}f is based on a dense array of spins in a grid of quantum dots. Individualised control of a subset of the qubits can be performed by reconfiguring the electrostatic potential and applying microwave excitations, both achieved directly by the top gate that defines the quantum dot. This on-demand activation of spin-orbit effects would significantly simplify the design and operation of large scale quantum processors by removing the need for additional complex nanomagnet or antennae arrays~\cite{boter2021spiderweb, li2018crossbar}. 

\begin{figure*}[ht]
    \centering
    \includegraphics[width = 1\textwidth, angle = 0]{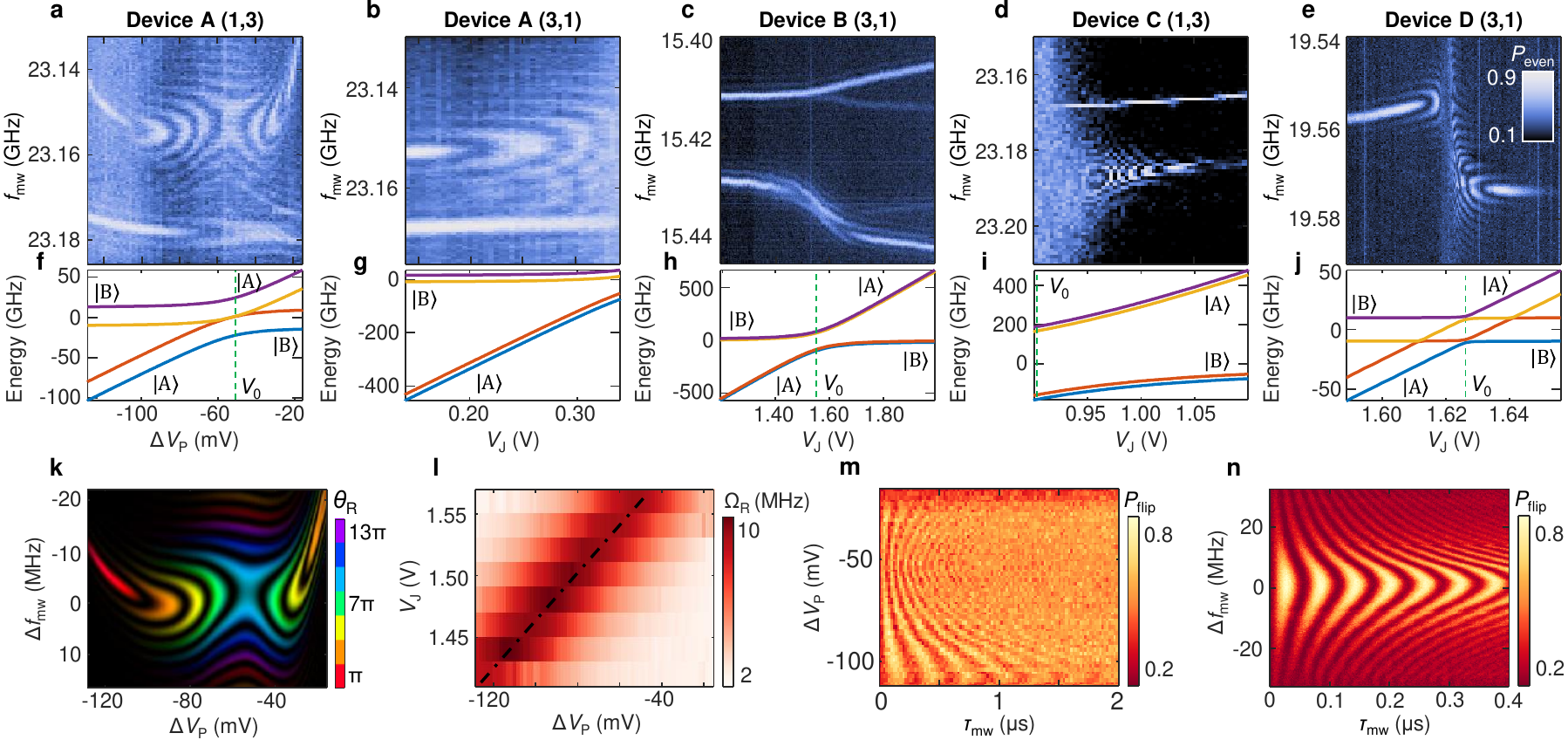}
    \caption{\textbf{Pulsed Electron Spin Orbit Spectroscopy (PESOS).} 
    \textbf{a}, PESOS map, consisting of the probability of measuring a spin flip after a burst of microwave of fixed power and duration, while varying the microwave frequency and a gate electrode voltage bias. The power and duration of the burst are roughly calibrated to correspond to a $\pi$-rotation for a pure spin state. Fringes appear as a function of gate voltage as the spin-photon coupling becomes more intense, and the spin being rotated by several $\pi$.
    \textbf{b-e}, Additional PESOS maps measured in different devices, varying materials, charge configurations, magnetic fields and others \ifsupp (see details in Extended Data). \fi
    \textbf{f-j}, Four-level models that best reproduce the data from \textbf{a-e}, showing the variety of crossing regimes and their impact on the spin dynamics.
    \textbf{k}, Simulated PESOS map obtained by modelling the spin-orbit qubit as a two-level system with voltage-dependent Rabi and Larmor frequencies to fit the measured PESOS map \ifsupp (see Supp. Info.)\fi. The colours correspond to the angle of spin rotations. 
    \textbf{l}, Rabi frequencies extracted from the fitted simulations. 
    \textbf{m}, Measured Rabi oscillations over time as a function of $\Delta V_{\rm P}$, confirming the interpretation of the Rabi speed-up. 
    \textbf{n}, Measured Rabi chevron from device D, where the qubit is driven all-electrically via the gate CB1.}
    \label{fig:pesos}
\end{figure*}

\subsection{Pulsed electron spin-orbital spectroscopy}
While the orbital spectroscopy technique presented in \autoref{fig:sem}c is useful in narrowing the search range for a degeneracy point, ultimately it is the change in spin dynamics that will be the most reliable signature of the successful formation of a hybrid wavefunction. We show as a dashed black line in \autoref{fig:sem}c the trend of points that are identified as having maximum spin-orbit driving. This identification is obtained by a technique we call pulsed electron spin-orbital spectroscopy (PESOS). It consists of applying a microwave pulse of fixed duration and power and measuring its effect on the spins as a function of the microwave frequency and gate voltages. Optimal visibility of the spin resonant frequency is obtained when the pulse duration and amplitude match with the condition for a spin flip.

\autoref{fig:pesos}a-e show examples of PESOS maps generated by double quantum dots in which spins are initialised and measured using parity readout~\cite{yang2019silicon,seedhouse2021pauli}. Where two peaks are observed at different frequencies, these correspond to spins in each of the two dots. All maps, taken from different devices and charge configurations, have identifiable hybridization points where the probability of a spin flip forms oscillations in at least one of the resonance lines. These oscillations are the result of an enhancement of the efficiency of the EDSR, which results in multiple spin flips with the same microwave power. Simulations of gate-dependent Rabi and Larmor frequencies, shown for example in \autoref{fig:pesos}k for the measurement from \autoref{fig:pesos}a, can be used to extract the magnitude of the speed-up and help interpret the PESOS maps. 

By measuring PESOS maps for different biases $V_{\rm J}$ and $\Delta V_{\rm P}$, we can extract the bias configuration that provides the largest speed-up in Rabi frequency, as shown in \autoref{fig:pesos}l. This allows us to completely reconstruct the line in the charge stability diagram in \autoref{fig:sem}c that corresponds to a hybrid ground state. \autoref{fig:pesos}m shows the complete Rabi oscillations of the spin, confirming our interpretation of the PESOS maps. We also use this interpretation to guide the experimental search for the degeneracy point shown in \autoref{fig:pesos}b \ifsupp (see also Extended Data)\fi. The regularity with which we find these hybridisation points is encouraging for the prospects of scalability of this technology. \autoref{fig:pesos}a and b were taken using different charge configurations in the same device. \autoref{fig:pesos}c-e are three other devices, with different operation modes, material stacks and microwave excitation strategies, measured in two different cryogenic setups. \ifsupp Details of the differences between devices A, B, C and D are given in the Supplemental Material.\fi

The taxonomy of the spin-orbit effects in \autoref{fig:pesos}a-e is related to the particularities of the orbital states A and B in each of the devices and dot configurations. The hybridization may involve states with different valley configurations (under a rough interface)~\cite{hao2014electron, bourdet2018all, corna2018electrically}, in-plane orbitals~\cite{kyriakidis2007universal} or even with interaction-induced charge transitions such as in Wigner molecules~\cite{ercan2021strong,abadillo2021wigner}. For each of these transitions, the hybridization energy gap compares differently to the spin splitting energy, leading to significant qualitative differences as can be seen in the eigenvalues of the fitted four-level models, shown in \autoref{fig:pesos}f-j.

We note that device D is driven all-electrically by applying a microwave field directly to the CB1 gate (see \autoref{fig:sem}a), while devices A to C are driven by the coplanar waveguide antenna, which creates both electric and magnetic fields~\cite{dehollain2012nanoscale}.

\begin{figure}
    \centering
    \hspace{-0.6cm}
    \includegraphics[width = \columnwidth, angle = 0]{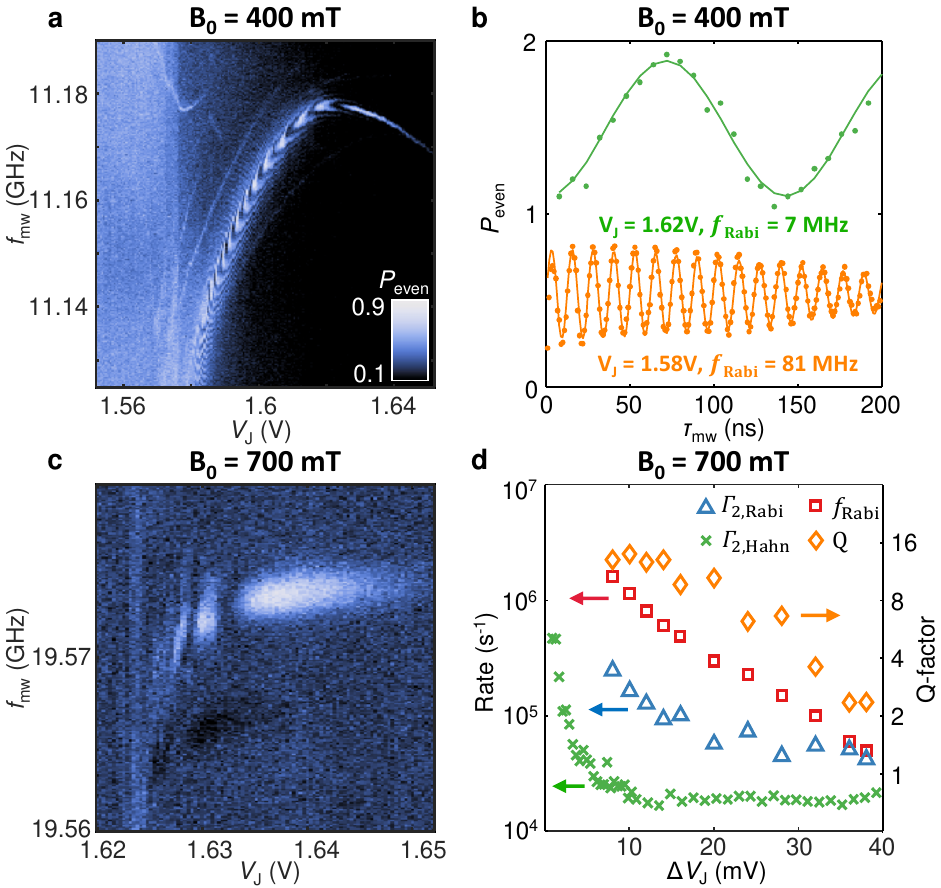}
    \caption{\textbf{Impact of Zeeman energy and detuning on qubit dynamics.} 
    \textbf{a}, PESOS map of device D at an external magnetic field of 400~mT measured with square pulses. At 400~mT the resonance can be tracked until very near the degeneracy point, resulting in an increase in Rabi frequency of almost three orders of magnitude.
    \textbf{b}, Rabi measurements at $V_{\rm J} = 1.62$~V resulting in 7~MHz oscillations (offset for clarity) and $V_{\rm J}=1.58$~V resulting in 81~MHz oscillations. (The Rabi frequency becomes immeasurably low for $V_{\rm J}>1.65$~V).
    \textbf{c}, PESOS map of device D at 700~mT measured with Gaussian pulses. At 700~mT the Rabi speed-up is more modest, but the qubit frequency is also less strikingly affected by electric field fluctuations, resulting in an overall more precise qubit operation compared to 400~mT. 
    \textbf{d}, Rabi frequency (red squares), Rabi decay rate (blue triangles), and Q-factor (orange diamonds) increase with proximity to the degeneracy point, with Q-factor reaching an optimal value at ${\Delta}V_{\rm J} = 13.6$~mV. The Hahn echo \ifsupp (see Extended Data) \fi coherence decay rate (green crosses) increases only very close to the degeneracy point.}
    \label{fig:zeeman}
\end{figure}

\subsection{Qubit performance}
We turn our attention to the tunability of the hybridization characteristics, focusing on the device from \autoref{fig:pesos}e (namely, device D). This is the device with the most marked effects of the orbital degeneracy on spins among the devices studied here. We measure additional PESOS maps at two different magnetic fields, $B_0=400$~mT in \autoref{fig:zeeman}a and $B_0=700$~mT in \autoref{fig:zeeman}c (the latter is the same field as \autoref{fig:pesos}e, but adopting Gaussian pulses and focused near the degeneracy point). At 400~mT, we observe the largest enhancement in Rabi frequency across all experiments (see \autoref{fig:zeeman}a). Reaching $f_{\rm Rabi}=81$~MHz close to the degeneracy point, we achieve the fastest $\pi/2$ qubit rotation in 3~ns (\autoref{fig:zeeman}b, lower curve). The spin-orbit interaction continues to decrease for increasing $V_{\rm J}$, leading to Rabi frequencies $<125$~kHz at $V_{\rm J}> 1.65$~V (data not shown and bias point out of range in \autoref{fig:zeeman}a), recovering the regime of vanishingly small spin-orbit interactions. For comparison, we also show the measurement at the point where the Larmor frequency is in first order insensitive to noise in the J gate voltage ($V_{\rm J} = 1.62$~V), in which the Rabi frequency is $f_{\rm Rabi}=7$~MHz (\autoref{fig:zeeman}b, upper curve).

At 700~mT the qubit states become more convoluted, with poorer initialisation fidelity and the appearance of additional transitions, which pollute the two-level nature of the system. Indeed, the small hybridization gap extracted in \autoref{fig:pesos}j is indicative of susceptibility to the appearance of undesirable diabatic transitions and leakage to the excited orbital. However, with careful initialisation strategies and using Gaussian pulses to avoid the leakage of the qubit into undesired excited states, it is possible to achieve PESOS maps with good visibility (\autoref{fig:zeeman}c) and coherent driving (\autoref{fig:zeeman}d). 

In \autoref{fig:zeeman}d we characterise the impact of the orbital hybridization on the coherent spin driving. Near the degeneracy point $\Delta V_{\rm J}=0$~V, the Stark shift ${\rm d}f/{\rm d}V_{\rm J}$ becomes very large, leading to a faster damping rate of the Rabi oscillations $\Gamma_{2,{\rm Rabi}}$. However, the Rabi frequency improvement outpaces the decoherence amplification, resulting in a higher Q-factor $=2{\times}f_{\rm Rabi}/{\it \Gamma}_{2,{\rm Rabi}}$ close to the degeneracy point. 

We devise and perform a Hahn echo experiment where for the same microwave control point ($\Delta V_{\rm J} = 13.6$~mV), the idle wait times are offset by $\Delta V_{\rm J}$ to analyse the potential for performing fast control near the spin-orbit mixing point $\Delta V_{\rm J}=0$~V, while idling at a bias point where the spin is decoupled from its orbital motion at large $\Delta V_{\rm J}$ (see \autoref{fig:zeeman}d). Interestingly, for $\Delta V_{\rm J}>10$~mV the Hahn echo coherence time saturates, indicating that the residual spin-orbit coupling does not introduce any additional decoherence. As this is still a point of strong Rabi enhancement, it is possible to operate this particular qubit in a configuration that enjoys both fast control and long coherence times, granted that some dynamical decoupling steps are incorporated in the qubit operation. 

\begin{figure}
    \centering
    \includegraphics[width = \columnwidth, angle = 0]{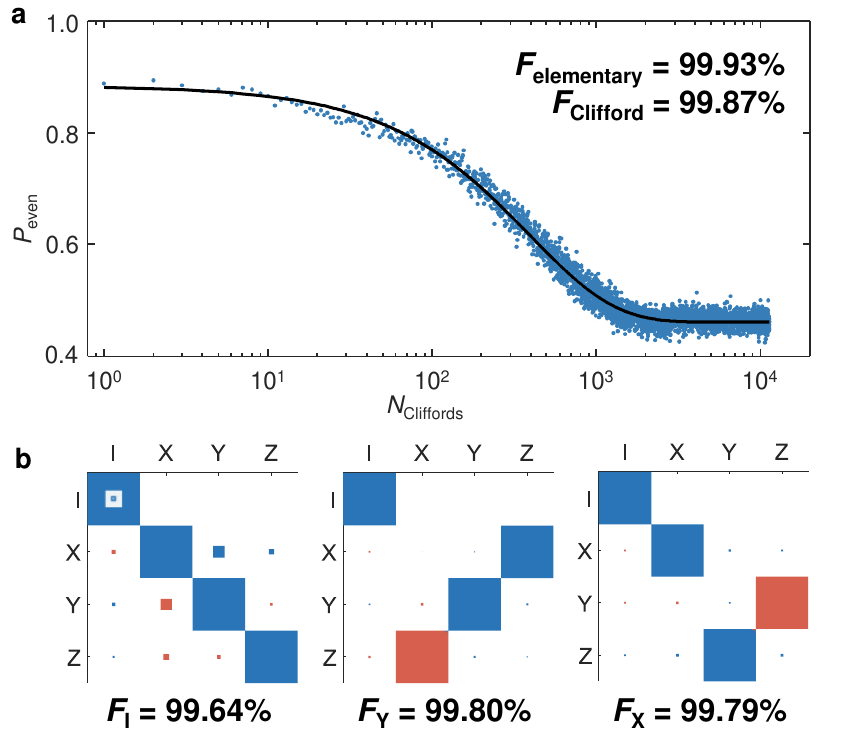}
    \caption{\textbf{Qubit fidelity and errors.}
    The device is configured to achieve a high-quality single qubit and benchmarked using standard techniques.
    \textbf{a}, Measurement of single qubit gate fidelity via randomised benchmarking on the Clifford set. The average elementary gate and average Clifford gate have $99.93$\% and $99.87$\% fidelity, respectively.
    \textbf{b}, The fidelity and errors associated with the individual elementary gates are characterised with gate set tomography. Projected populations are displayed as areas, with blue and red representing positive and negative polarity. Concentric squares in the top-left position of the first panel represent populations of 1.0, 0.1, 0.01, and 0.001.}
    \label{fig:qubit}
\end{figure}


For the remaining experiment, we tune device D to an alternative bias point in the (1,3) charge configuration and apply the microwave drive to the J gate to probe spin-orbit coupling in line with the array of dots. This configuration yields a degeneracy point with strong spin-orbit coupling in a region with only a weak voltage dependence on the qubit Larmor and Rabi frequencies (see Extended Data), making the qubit substantially more resistant to noise. The single qubit gate fidelity is assessed in this configuration via randomised benchmarking on the Clifford set, achieving a 99.93\% elementary gate fidelity as shown in \autoref{fig:qubit}a, well above the threshold for error corrected fault tolerance \cite{fowler2012surface}. Additionally we assess the fidelity of the individual elementary gates using Gate Set Tomography~\cite{nielsen2020gate} shown in \autoref{fig:qubit}b resulting in average fidelities of 99.64\%, 99.79\% and 99.80\% for the I, X, and Y gate, respectively. 





\subsection{Outlook}
Entering a new realm of ultra-fast single spin control creates new questions regarding the physics of these systems and their application for quantum information processing. For example, the dominant source of control errors for this resonance method are unknown. Pulse engineering for magnetically driven spin qubits in similar devices have led to significant improvements in control fidelity, achieving error rates below 0.05\%~\cite{yang2019silicon}. However, these strategies can only be translated to the electric driving approach discussed here once the sources of error are well understood and characterised. 

Another question left open in our analysis is in regard of the controllability of the hybridization gap. Comparison of the spin-orbit effect in all four devices investigated here creates confidence on the ubiquity of this phenomenon. Hence, its applicability as the main control strategy for qubit devices depends on the regularity of the resulting EDSR speed-up and dependability on being able to shape the orbital hybridization gap in accordance with one's needs. 

Switchability of the electrical dependency allows us to tackle one of the most cumbersome aspects of quantum information, which is that the addressability of a qubit often must be traded off against its noise resilience. This allows us to turn on the degeneracy for control and turn it off to harvest the long intrinsic coherence of Si qubits while idling. This prospect comes with the cost of an additional characterisation step for the quantum processor to achieve this degeneracy. Hence, from the scalability perspective it is crucial to further understand how to achieve this degeneracy in a consistent way in a given dot. We believe that the consistency of achieving this degeneracy in four different devices already in this first demonstration gives confidence that a consistent method is achievable and we anticipate more experiments and theoretical work in this direction in future. 

We emphasise that enhanced electric driving of the spin is merely one consequence of the ability to controllably create hybridised wavefunctions with coherent spin states. Extensions of this result could lead to strategies to couple spins to photons~\cite{yao2005theory,mi2018coherent}, as well as lead to long-range two-qubit gates via spin-dependent electric dipolar coupling, similar to strategies such as the Rydberg gates~\cite{jaksch2000fast,crane2021rydberg} and M\o lmer-S\o rensen gates~\cite{sorensen1999quantum}, previously demonstrated in atomic qubits, or predicted for electron-nuclear flip-flop qubits in silicon~\cite{tosi2017Silicon}.

\section{Methods}

\subsection{Measurement Setup}
Devices A \& C were measured in an Oxford Kelvinox 400HA dilution refrigerator. DC bias voltages are generated from Stanford Research Systems SIM928 Isolated Voltage Sources. Gate pulse waveforms are generated from a Tektronix AWG5208 Arbitrary Waveform Generator (AWG) and combined with DC biases using custom linear bias-tees. 

Devices B \& D were measured in a Bluefors XLD400 dilution refrigerator. DC bias voltages are generated from Basel Precision Instruments SP927 DACs. Gate pulse waveforms are generated from a Quantum Machines OPX and combined with DC biases using custom linear bias-tees.

The SET current of devices A, B, \& C are amplified using a room temperature I/V converter (Basel SP983c) and sampled by a digitiser (Gage Octopus CS8389 for devices A \& B, QM OPX for device C). The SET of device D is connected to a tank circuit and measured via reflectometry, where the source tone is generated from the QM OPX, and the return signal amplified with a Cosmic Microwave Technology CITFL1 LNA at the 4K stage, and a Mini-circuits ZX60-P33ULN+ and Mini-circuits ZFL-1000LN+ at room temperature, before being digitised and demodulated with the QM OPX.

For all devices, microwave pulses are generated from a Keysight PSG8267D Vector Signal Generator, with I/Q and pulse modulation waveforms generated from the AWGs.

\subsection{Theoretical Modeling and Fits}
Here, we summarise the theoretical method involved in obtaining the four-level energy diagrams shown in Fig.~\ref{fig:pesos}f-j. The goal of this method is to obtain a description of our system based on the PESOS maps using an effective four-level model, consisting of two spin-$\frac{1}{2}$ systems, $\mathrm{A}$ and $\mathrm{B}$. These quantum states can be either valley or orbital states, depending on the specific system, and are represented by $\ket{\mathrm{A}}$ and $\ket{\mathrm{B}}$. The spin states are split by the Zeeman splitting in the presence of a magnetic field and thus, forming a total of four non-degenerate states. \ifsupp A full description of the Hamiltonian is contained in the Supplementary Information.\fi

The four-level model is fitted to two different sets of information, one is the qubit frequency $f_0$, and the other is the Rabi frequency given by $f_\mathrm{Rabi}$. To obtain these information from the PESOS maps as shown in Fig.~\ref{fig:pesos}a-e, we extract vertical line traces of $P_\mathrm{even}$ as a function of the driving frequency $f_\mathrm{mw}$ at each voltage value. We can then fit these traces to the Rabi equation given by: 
\begin{align}
    P_\mathrm{even} = \frac{A f_\mathrm{Rabi}^2 \left[1-\cos\left(\tau\sqrt{f_\mathrm{Rabi}^2+(f_\mathrm{mw}-f_0)^2}\right)\right]}{f_\mathrm{Rabi}^2+(f_\mathrm{mw}-f_0)^2}+\delta{A}
\end{align}
where $A$ is the amplitude of the oscillations, $f_\mathrm{Rabi}$ is the Rabi frequency, $f_\mathrm{mw}$ is the driving frequency, $f_0$ is the resonant qubit frequency, $\tau$ is the total time of the driving pulse, and $\delta{A}$ is amplitude offset of the oscillations. From this fit, we can extract both the Rabi frequency $f_\mathrm{Rabi}$, and the qubit frequency, given by $f_0$, as a function of the gate voltage (either $V_\mathrm{J}$ or $\Delta V_\mathrm{P}$). With these information, we are also able to obtain the simulated PESOS maps as shown in Fig.~\ref{fig:pesos}k by plotting the Rabi equation for each voltage value with the fitted parameters.

These extracted values of $f_\mathrm{Rabi}$ and $f_0$ as a function of gate voltage will be the target fit values of the four-level model. By varying the parameters of the four-level model Hamiltonian, we perform a non-linear least squares fit of both the Rabi frequencies $f_\mathrm{Rabi}$ and the qubit frequency $f_0$ simultaneously, minimizing the difference between the calculated values from the four-level model and target values obtained from fitting to the Rabi equation. The output of this least squares fit are the Hamiltonian parameters describing the system. \ifsupp More details on the fitting procedure and fitted values are contained in the Supplementary Information. \fi Finally, these fitted parameters will enable us to calculate the eigen-energies of the Hamiltonian and obtain the energy diagrams as shown in Fig.~\ref{fig:pesos}f-j.

\subsection{Randomised Benchmarking}
The benchmarking sequences used in \autoref{fig:qubit}a are constructed from elementary $\pi/2$ gates {X, Y, -X, -Y}, $\pi$ gates [X,X], [Y,Y], and an I-gate which is implemented as a sequence of [X,X,-X,-X,-X,-X,X,X]. Each Clifford gate contains on average 1.875 elementary gates.

For each data point, an average probability is taken from 20 randomised sequences, each averaged over 100 shots. Due to a hardware memory limit, sequences longer than 1420 Cliffords are executed as repetitions of half, quarter, or eighth-length sequences as necessary. Repeated sections have a minimum length of 710 Cliffords.

\section{Data Availability}
The datasets generated and/or analysed during this study are available from the corresponding authors on reasonable request.

\section{Code Availability}
The analysis codes that support the findings of the study are available from the corresponding authors on reasonable request.

\section{Acknowledgements}
We acknowledge helpful conversations and technical support from A. Dickie. We acknowledge support from the Australian Research Council (FL190100167 and CE170100012), the US Army Research Office (W911NF-17-1-0198), and the NSW Node of the Australian National Fabrication Facility. The views and conclusions contained in this document are those of the authors and should not be interpreted as representing the official policies, either expressed or implied, of the Army Research Office or the US Government. The US Government is authorized to reproduce and distribute reprints for Government purposes notwithstanding any copyright notation herein. W.G., M.F., J.Y.H., J.D.C., and S.S. acknowledge support from Sydney Quantum Academy.

\section{Author information}
\subsection{Author Contributions}

T.T. measured devices A and C first observing the enhanced SOI Rabi in A.
W.G. measured devices B and D demonstrating the EDSR without a micromagnet in D.
J.Y.H participated in qubit benchmarking in D, with M.F \& W.G. participating in the analysis.
W.H.L participated in the measurements with all devices.
Experiments were done under A.L., A.S., A.S.D., and C.H.Y.'s supervision.
W.H.L. and F.E.H. fabricated the devices, with A.S.D.'s supervision on enriched $^{28}$Si wafers supplied by K.M.I., N.V.A., H.-J.P., and M.L.W.T. 
S.S. designed the RF setup for devices B and D.
A.S., C.H.Y., and R.C.C.L. developed control strategies for enhanced EDSR.
P.Y.M., M.F., and J.D.C. developed models of spin-orbital degeneracy, with A.S. and C.E.'s supervision.
W.G., A.S., M.F., T.T., W.H.L., J.Y.H, F.E.H., A.L., C.H.Y., A.M., and A.S.D. wrote the manuscript, with input from all co-authors.

\subsection{Corresponding Authors}
Correspondence to W. Gilbert, A. Saraiva, or A. S. Dzurak.

\section{Competing Interests}
T.T., W.H.L., R.C.C.L., A.L., C.H.Y., A.S. and A.S.D. are inventors on a patent related to this work (AU provisional application 2021901923) filed by the University of New South Wales with a priority date of 25th June 2021. All other authors declare they have no competing interests.

\bibliographystyle{naturemag}
\bibliography{EDSR}

\ifsupp

\input{extData}
\fi

\end{document}


\title{Supplementary Information: On-demand electric control of spin qubits}
	
\author{Will Gilbert}
\altaffiliation[]{These co-authors had equal contributions to this work.}
\author{Tuomo Tanttu}
\altaffiliation[]{These co-authors had equal contributions to this work.}
\author{Wee Han Lim}
\author{MengKe Feng}
\author{Jonathan Y. Huang}
\author{Jesus D. Cifuentes}
\author{Santiago Serrano}
\author{Philip Y. Mai}
\author{Ross C. C. Leon}
\author{Chris C. Escott}
\affiliation{School of Electrical Engineering and Telecommunications, The University of New South Wales, Sydney, NSW 2052, Australia}
\author{Kohei M. Itoh}
\affiliation{School of Fundamental Science and Technology, Keio University, Yokohama, Japan}
\author{Nikolay V. Abrosimov}
\affiliation{Leibniz-Institut für Kristallzüchtung, 12489 Berlin, Germany}
\author{Hans-Joachim Pohl}
\affiliation{VITCON Projectconsult GmbH, 07745 Jena, Germany}
\author{Michael L. W. Thewalt}
\affiliation{Department of Physics, Simon Fraser University, British Columbia V5A 1S6, Canada}
\author{Fay E. Hudson}
\author{Andrea Morello}
\author{Arne Laucht} 
\author{Chih Hwan Yang}
\author{Andre Saraiva}
\author{Andrew S. Dzurak}
\affiliation{School of Electrical Engineering and Telecommunications, The University of New South Wales, Sydney, NSW 2052, Australia}
\date{\today}

\maketitle


\tableofcontents

\newpage

\section*{Supplementary Discussion: Origin of spin-orbit parameters and implications for spin driving}

In this supplementary section, we explain the analysis of the experimental results and the construction of the four-level model that can be used to describe the transition in qubit frequency between orbitals as well as the speed up in Rabi frequency near the degeneracy points. This modeling protocol will finally allow us to obtain the energy diagrams shown in Fig.~2f-j in the main text.

\subsection{Analysis of PESOS maps}
\label{subsec:pesos}
We begin by extracting the key quantities from the PESOS maps shown in Fig.~2 of the main text, which are the qubit frequency, denoted as $f_\mathrm{0}$, and the Rabi frequency, denoted as $f_\mathrm{Rabi}$. 

\begin{figure}
    \centering
    \includegraphics[width = \textwidth, angle = 0]{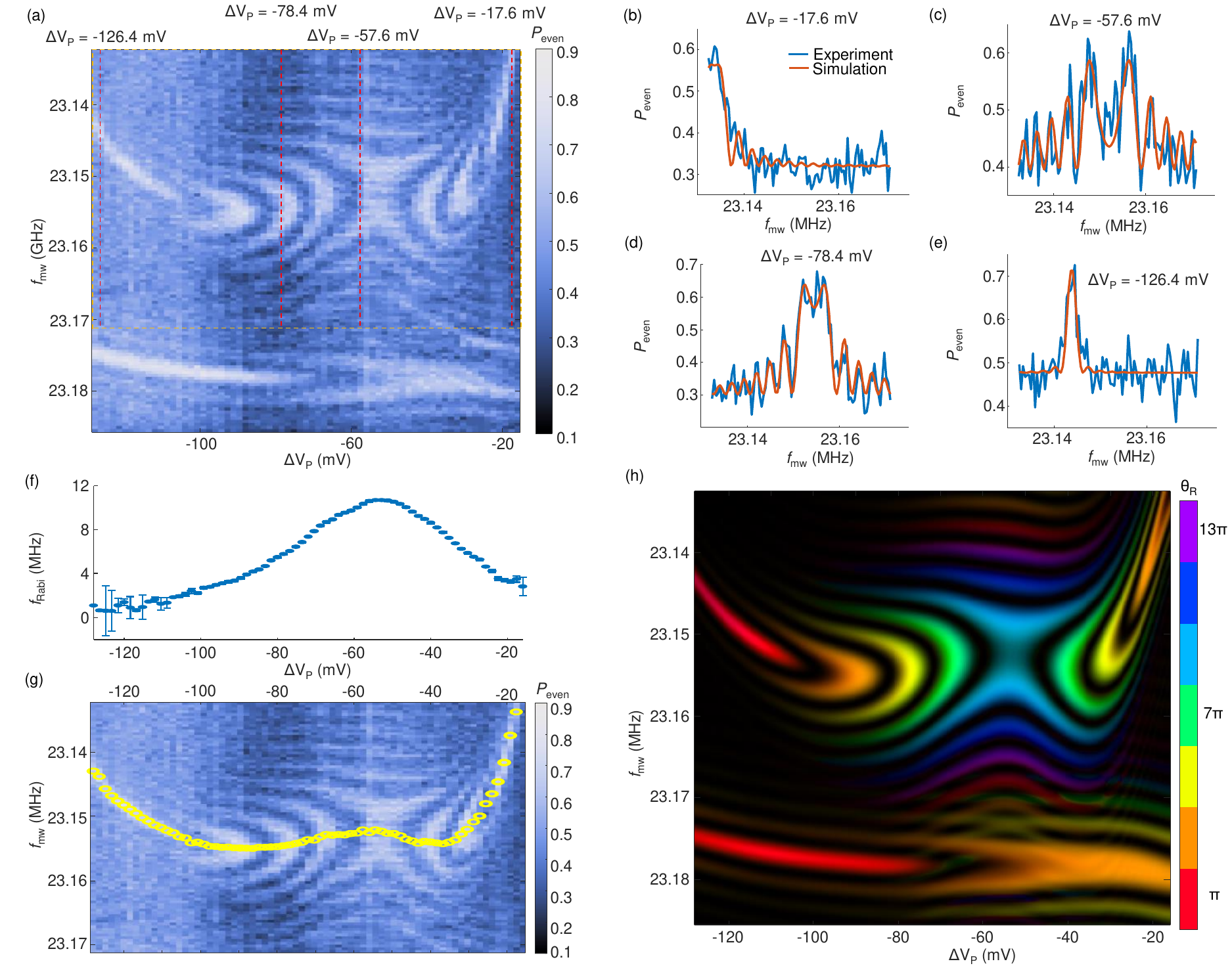}
    \caption{\textbf{Rabi frequency fitting.}
    \textbf{a}, PESOS map for Device A in the (1,3) configuration. The red dotted lines indicate examples of line cuts. We focus only on the upper transition (outlined by the yellow dotted lines), with a clear speed up in Rabi frequency as well as a transition in qubit frequency.
    \textbf{b-e}, Line cuts of $P_\mathrm{even}$ oscillations taken at the indicated dotted lines in \textbf{a}. Blue traces indicate the experimental data and red traces are the fitted lines based on Eq.~\ref{eqn:Rabi}.
    \textbf{f}, Extracted Rabi frequency $f_\mathrm{Rabi}$ as a function of voltage $\Delta V_\mathrm{P}$. Error bars of the fit are also shown.
    \textbf{g}, Fitted qubit dispersion $f_0$ superimposed onto a PESOS map.
    \textbf{h}, Simulated PESOS map based on fitted traces shown in \textbf{b-e}.}
    \label{fig:supp1}
\end{figure}

We firstly identify which qubit is suspected to have an orbital transition by inspecting qualitative signatures in its PESOS map. The qubit of interest will exhibit both a speed-up in Rabi frequency $f_\mathrm{Rabi}$ indicated by oscillation patterns for fixed pulse time, and a transition in the qubit frequencies $f_0$ indicated by a non-linear change in the resonance (non-linear Stark shift). The upper transition in the PESOS map shown in Supp Fig.~\ref{fig:supp1}a is an example. We then take line cuts along the frequency axis ($f_\mathrm{mw}$) at separate voltage values ($\Delta V_\mathrm{P}$ in this case) with examples indicated in Supp Fig.~\ref{fig:supp1}a. These line traces are also plotted in blue in Supp Figs.~\ref{fig:supp1}(b-e). These traces are oscillations of $P_\mathrm{even}$ probabilities and we can fit them according to the following equation,
\begin{align}
    P_\mathrm{even} = \frac{A f_\mathrm{Rabi}^2 \left[1-\cos\left(\tau\sqrt{f_\mathrm{Rabi}^2+(f_\mathrm{mw}-f_0)^2}\right)\right]}{f_\mathrm{Rabi}^2+(f_\mathrm{mw}-f_0)^2}+\delta{A},
\end{align}
where $A$ is the amplitude of the oscillations, $f_\mathrm{Rabi}$ is the Rabi frequency, $f_\mathrm{mw}$ is the driving frequency from the microwave source, $f_0$ is the resonant frequency which can be interpreted as the qubit frequency, $\tau$ is the total time of the driving microwave pulse, and $\delta{A}$ is a constant shift in the amplitude of the probabilities due to SPAM errors. Fitted traces of $P_\mathrm{even}$ are plotted in red in Supp Figs.~\ref{fig:supp1}(b-e). 

The Rabi frequency $f_\mathrm{Rabi}$ is plotted in Supp Fig.~\ref{fig:supp1}f along with the fitting error bars. In Supp Fig.~\ref{fig:supp1}g, we superimposed the fitted qubit frequencies $f_0$ onto the PESOS map, indicating that the qubit frequencies correspond to the center of the oscillations, as expected, and also show that this empirical fitting protocol is of satisfactory accuracy. 

Finally, as a demonstration of the capabilities of the fitting protocol, we generate a simulated PESOS map colour coded such that the colours represent the number of $\pi$ rotations ($\theta_\mathrm{R}$) that the qubit undergoes. As the Rabi frequency speeds up and slows down again with voltage, the same fixed pulse duration causes more or less qubit rotations (with red representing one $\pi$ rotation and purple representing 13 $\pi$ rotations). 

\subsection{Four-level model of spin-orbital quasi-degeneracy}
\label{subsec:fourlevel}

We have described how we can obtain a set of parameters  for fitting the PESOS maps. From the theoretical point of view, we interpret these parameters based on properties of quasi-degenerate orbitals. Our goal is to find an effective model that captures the qualitative features of the qubit and Rabi frequencies. This will typically involve making conjectures about the degrees of freedom involved in the orbital excitation of a multielectron state in a quantum dot, estimating the underlying first-principles microscopic Hamiltonian, then extracting the effective two-level system subspace, by a Schrieffer-Wolff Transformation for example, that reduces to the phenomenological model.

For this purpose, the simplest model consists of two spin-$\frac{1}{2}$ states with orbital parts designated by $A$ and $B$, which represent the two different sets of states away from the degeneracy point, \textit{i.e.}, $\ket{A}, \ket{B}$ are the quantum states representing two different valley states, orbital states etc. The use of our model is agnostic to the exact nature of these states and therefore, we will refer to them generally as orbital states for simplicity. We also operate in electronic configurations with a single valence electron, and therefore our system can be described in the basis of $\{\ket{A},\ket{B}\}\otimes\{\ket{\uparrow},\ket{\downarrow}\}$. Here, the spin degree of freedom is in general a pseudo-spin due to spin-orbit coupling. The $g$-factors for both orbitals are typically  close to the bulk value in silicon, with small variations of $\approx$0.1\% from qubit to qubit and between orbitals due to surface roughness. This implies that the Zeeman splitting $E_\mathrm{Z}^\mathrm{A}$ will in general be different from $E_\mathrm{Z}^\mathrm{B}$.

The general form of the desired Hamiltonian is
\begin{align}
    H_0 = \begin{pmatrix}
        H_\mathrm{A} & H_\mathrm{c} \\ H_\mathrm{c}^\dagger & H_\mathrm{B}
    \end{pmatrix}\;,
    \label{eqn:ham}
\end{align}
where $H_\mathrm{A}$ and $H_\mathrm{B}$ describe quantum subsystems $\mathrm{A}$ and $\mathrm{B}$, respectively, far from any degeneracy, and $H_\mathrm{c}$ describes the coupling between subsystems $\mathrm{A}$ and $\mathrm{B}$. Each of these terms ($H_\mathrm{A}$, $H_\mathrm{B}$, and $H_\mathrm{c}$) are $2\times2$ blocks. For subsystem $\mathrm{A}$,
\begin{align*}
    H_\mathrm{A} = \left[\frac{1}{2}E_\mathrm{Z}^\mathrm{A} + \eta_\mathrm{A}\left(V_\mathrm{G}-V_0\right) \right]\sigma_z\;,
\end{align*}
where $E_\mathrm{Z}^\mathrm{A}$ is the Zeeman energy of subsystem $\mathrm{A}$ and $\eta_\mathrm{A}$ is the linear part of its Stark shift (all the non-linearity of the Stark shift in our model stems from the resulting orbital hybridization near the degeneracy point), $V_\mathrm{G}$ is the gate voltage, and $\sigma_z$ is the Pauli $z$ operator acting on the spin basis $\{\ket{\uparrow}, \ket{\downarrow}\}$. The definition of the reference voltage $V_0$ is discussed next. For subsystem $\mathrm{B}$,
\begin{align*}
    H_\mathrm{B} = \left[\frac{1}{2}E_\mathrm{Z}^\mathrm{B} + \eta_\mathrm{B}\left(V_\mathrm{G}-V_0\right) \right]\sigma_z + \alpha_\mathrm{rel}\left(V_\mathrm{G}-V_0\right)\mathds{1}\;,
\end{align*}
where the Hamiltonian takes on a similar form to that for subsystem $\mathrm{A}$, except with the corresponding parameters for $\mathrm{B}$ and the additional term describing the effect of the gate voltage on the energy separation between orbitals.

We model the effect of a gate with voltage $V_\mathrm{G}$ bringing $\ket{\mathrm{B}}$ into alignment with $\ket{\mathrm{A}}$ by adding the term $\alpha_\mathrm{rel} (V_\mathrm{G} - V_0)\mathds{1}$ to $H_\mathrm{B}$ where $\alpha_\mathrm{rel}$ is the differential lever arm between $\ket{\mathrm{A}}$ and $\ket{\mathrm{B}}$. Here, the definition of $V_0$ becomes clear -- it is the voltage bias at which the two states A and B would be degenerate (which gets slightly shifted in the presence of a difference in Zeeman splitting). Finally, the last term in Eq.~\ref{eqn:ham} is one that describes the coupling between $\ket{\mathrm{A}}$ and $\ket{\mathrm{B}}$, which sets the energy gap at the anticrossing,
\begin{align*}
    H_\mathrm{c} = \begin{pmatrix}
        \Delta + \Delta_\mathrm{sd} & \Delta_\mathrm{sf} \\
        \Delta_\mathrm{sf} & \Delta - \Delta_\mathrm{sd}
    \end{pmatrix}\;,
\end{align*}
Here, $\Delta$ is the spin-independent coupling rate between $\ket{\mathrm{A}}$ and $\ket{\mathrm{B}}$. The presence of spin-prbit coupling creates a spin-dependent coupling term $\Delta_\mathrm{sd}$, as well as a spin-flip coupling $\Delta_\mathrm{sf}$. The existence of both spin-conserving and spin-flip terms stem from the non-alignment of the qubit quantization axis with the crystallographic axes of the device, as well as the reduced symmetry due to interface roughness. This will be shown explicitly in a later section. Note that a $\sigma_y$ term in $H_\mathrm{c}$ is not necessary for fitting the qubit frequency because it only produces a phase shift in the transverse coupling - but it may be important for fitting Rabi frequencies as will be explained in subsection~\ref{subsec:socorigin}. 

Therefore, the full Hamiltonian in the basis of $\{\ket{\mathrm{A}},\ket{\mathrm{B}}\}\otimes\{\ket{\uparrow},\ket{\downarrow}\}$ is 
\begin{equation}
    H_0 = 
        \begin{pmatrix}
            \frac{1}{2}E_\mathrm{Z}^\mathrm{A} + \eta_\mathrm{A}\left(V_\mathrm{G}-V_0\right) & 0 & \Delta + \Delta_\mathrm{sd} & \Delta_\mathrm{sf} \\
            0 & -\frac{1}{2}E_\mathrm{Z}^\mathrm{A} - \eta_\mathrm{A}\left(V_\mathrm{G}-V_0\right) & \Delta_\mathrm{sf} & \Delta - \Delta_\mathrm{sd} \\
            \Delta + \Delta_\mathrm{sd} & \Delta_\mathrm{sf} & \frac{1}{2}E_\mathrm{Z}^\mathrm{B} + \left(\eta_\mathrm{B}+\alpha_\mathrm{rel}\right)\left(V_\mathrm{G}-V_0\right) & 0 \\
            \Delta_\mathrm{sf} & \Delta - \Delta_\mathrm{sd} & 0 & -\frac{1}{2}E_\mathrm{Z}^\mathrm{B} - \left(\eta_\mathrm{B}-\alpha_\mathrm{rel}\right)\left(V_\mathrm{G}-V_0\right).
        \end{pmatrix}
        \label{eqn:fullham}
\end{equation}

As we have described, this model works for arbitrary orbital states $\ket{\mathrm{A}}$ and $\ket{\mathrm{B}}$, and provides the theoretical basis for predicting both the change in qubit dispersion from one orbital to another, and the speed-up in Rabi frequency $f_\mathrm{Rabi}$ near the degeneracy point. These are the two key quantities that we will calculate using the model, and fit to the experimental results. 

To calculate the qubit dispersion, we first diagonalize the Hamiltonian numerically for a given set of parameters, which outputs the eigenenergies of the system. We then consider the energy difference between the lowest spin up and spin down states ($f_\mathrm{0} = E_\uparrow - E_\downarrow$), which is also the first excitation energy, as the qubit frequency. Generally for a large orbital coupling (for example in the cases of Supp Fig.~\ref{fig:supp2}l-n), the ground and excited orbital energies are well separated and the qubit frequency is also the difference in energy between the two lowest energy levels. In the case where the orbital coupling $\Delta$ is smaller than the Zeeman energies $E_Z^A$ and $E_Z^B$, there are multiple degeneracy points (for example in Supp Fig.~\ref{fig:supp2}o) and there is a need to track the lowest energy spin up state which changes in its ordering with detuning.

The second fitting parameter is the Rabi frequency, which is a function of detuning and is determined by,
\begin{align}
    f_\mathrm{Rabi} = \left|\bra{e}H_\mathrm{AC}\ket{g}\right|
\label{eqn:Rabi}
\end{align}
where $\ket{g}$ and $\ket{e}$ are respectively the ground and excited states. Again, in the case where the orbital coupling is lesser than the Zeeman energy (Supp Fig.~\ref{fig:supp2}o), it becomes necessary to track the states so that the Rabi frequency is always calculated between spin up and down states. The driving Hamiltonian, $H_\mathrm{AC}$, is defined as
\begin{align}
    H_\mathrm{AC} = \Omega_\mathrm{AC}\cdot\sigma_x\otimes\sigma_x \;,
\end{align}
where $\Omega_\mathrm{AC}$ is the magnitude of the spin-flip modulation resulting from the driving microwave pulse on the qubit. 

\subsection{Fitting Procedure}
\label{subsec:fitting}

Now, we have all the ingredients necessary to perform the fit, and we perform a non-linear least-squares fit of both the Rabi frequencies and the qubit dispersion simultaneously. The cost function in this minimization procedure is defined as
\begin{align}
    f_\mathrm{cost} = \sum (f_\mathrm{model} - f_\mathrm{PESOS})^2,
    \label{eqn:fcost}
\end{align}
where $f_\mathrm{model}$ and $f_\mathrm{PESOS}$ refer to the quantities obtained from the four-level model and extracted from the PESOS maps respectively. They include both the qubit dispersion $f_0$ and Rabi frequency quantities $f_\mathrm{Rabi}$. 

\begin{figure}
    \centering
    \includegraphics[width = \textwidth, angle = 0]{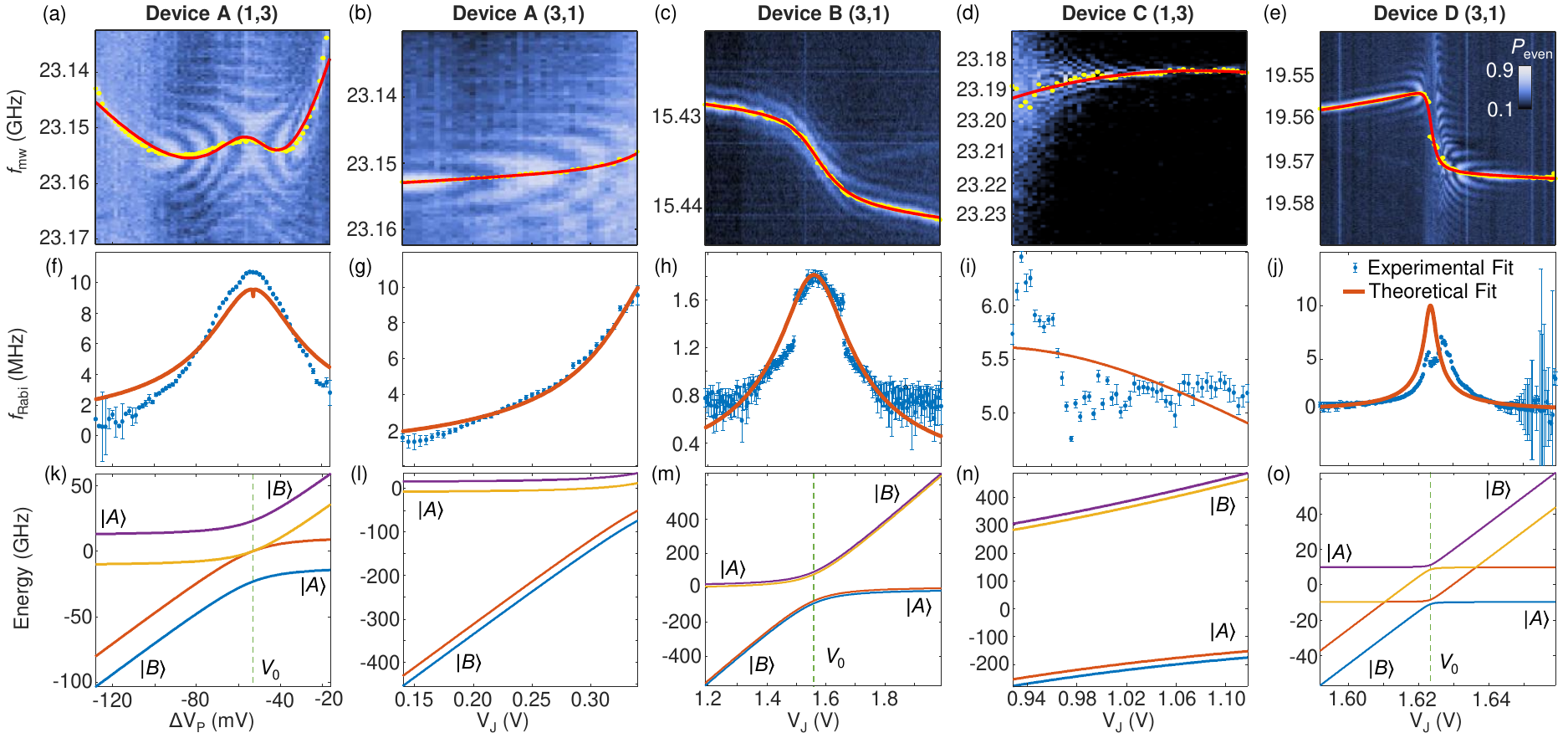}
    \caption{\textbf{Four-level model fit.}
    \textbf{a-e}, PESOS maps with both the fitted center frequency $f_0$ (yellow dots) and the fitted qubit dispersion from the four-level model (red lines). We show here only the relevant qubit of study.
    \textbf{b-j}, The extracted Rabi frequencies $f_\mathrm{Rabi}$ from the fitting to the PESOS map and the fitted Rabi frequencies from the four-level model. We note that for \textbf{j}, the error bars extend below $f_\mathrm{Rabi} = -5~\mathrm{MHz}$ but are not shown here in order to highlight the actual range of Rabi frequencies. Both the positive and negative error bars are of the same magnitude.
    \textbf{k-o}, The resultant four-level energy diagram from the fitted parameters.}
    \label{fig:supp2}
\end{figure}

From this point, the fitting procedure works to vary the parameters of the Hamiltonian as defined in Eq.~\ref{eqn:fullham}, such that we minimize the cost function as shown in Eq.~\ref{eqn:fcost}. We summarize the protocol as follows,
\begin{enumerate}
    \item Begin with an initial guess of the Hamiltonian parameters. This is an educated guess based on what we know of the system and does not have to be strictly accurate. How we can begin to guess at these quantities will be explained in the following sections.
    \item Using the \textit{lsqcurvefit} function in MATLAB, we input the Hamiltonian parameters necessary for calculating $f_\mathrm{0}$ and $f_\mathrm{Rabi}$ from the model, which will be compared with the target values previously obtained from the fitting to the Rabi equation. We also set upper and lower bounds to the fitting parameters of the Hamiltonian within the orders of magnitude that are expected.
    \item The output from this fitting procedure are the parameters of the Hamiltonian. The fit results are summarized in Table~\ref{table:fittedvals}.
\end{enumerate}

The results from this fitting protocol are summarized in Supp Fig.~\ref{fig:supp2}.

We plot in Supp Figs.~\ref{fig:supp2}a-e the PESOS maps with both the target (yellow dots) and fitted (red line) $f_0$ superimposed onto the 2D map. We include here only the qubit of concern, more specifically, only the qubit that shows a significant speed-up in Rabi frequency. Similarly, in Supp Figs.~\ref{fig:supp2}f-j, we plot the Rabi frequencies within the same voltage range as the PESOS maps, with the blue dots representing the target values of Rabi frequency which are obtained in the same way as shown in Supp Fig.~\ref{fig:supp1}f, and the red line being the fitted Rabi frequencies from the four-level model. We note that there are large errors in the Rabi frequency fit for extreme values of $V_\mathrm{J}$ in Supp Fig.~\ref{fig:supp2}j due to a reduction in the visibility of the Rabi oscillation peaks associated to the reduction in coherence. Finally, using these fitted parameters, we can diagonalize the Hamiltonian and obtain the eigen-energies as a function of the voltage as shown in Supp Figs.~\ref{fig:supp2}(k-o), similarly to Fig.~2 in the main text. The anticrossing for the transition from one orbital to another is indicated as a green dashed line and labeled by $V_0$, except in the case of Supp Fig.~\ref{fig:supp2}l,n, where the anticrossing is out of the plotted range. The expected dominant state (either $\ket{\mathrm{A}}$ or $\ket{\mathrm{B}}$) is also labeled, with the energies of $\ket{\mathrm{B}}$ shown to be changing significantly with gate voltage.

In the rest of the section, we will explain in detail the parameters involved in the fitting. There are many parameters involved in the construction of the Hamiltonian as shown in Eq.~\ref{eqn:fullham}, and in order for accurate and efficient fitting, it is necessary for us to consider the real system and reduce the number of degrees of freedom that we have to deal with. Of the many parameters that define the Hamiltonian, there are two of them which we fix in all the devices, those are the differential lever arm, $\alpha_\mathrm{rel}$ and the anticrossing position in voltage, $V_0$. 

The position of the anticrossing can be read directly from the voltage position of the Rabi frequency peak assuming a singular peak, as shown in Supp Figs.~\ref{fig:supp2}f-j, and therefore can be readily confirmed without fitting. Test fits can also be performed to empirically guess the position of the anticrossing in cases where the anticrossing may not be obvious, like in Supp Figs.~\ref{fig:supp2}g,i. 

On the other hand, to determine the differential lever arm $\alpha_\mathrm{rel}$ requires time-consuming additional measurements, for example, excited state magnetospectroscopy (main text Fig.~1c), which was only performed for device A in the (1,3) charge configuration. For fitting purposes, we take all the $\alpha_\mathrm{rel}$ to be on the same order of magnitude as obtained in device A. It is not possible to independently obtain the differential lever arm, $\alpha_\mathrm{rel}$, and the inter-orbital coupling, $\Delta$, directly from Supp Figs~\ref{fig:supp2}a-e, only their ratio $\alpha_\mathrm{rel}/\Delta$.

The other fitting parameters comprise the inter-orbital coupling $\Delta$, the spin-orbit couplings $\Delta_\mathrm{sd}$ and $\Delta_\mathrm{sf}$, the constant offset in frequency $c$, the difference in qubit frequencies between orbitals $\Delta{E_\mathrm{Z}}$, the linear Stark shifts of each orbital $\eta_\mathrm{A/B}$, and finally the amplitude of the electric driving $\Omega_\mathrm{AC}$. We note that not all of these parameters will necessarily be used for fitting. In particular, the Stark shifts are generally only used where we can observe the linear Stark shift regions in the PESOS maps. For example, in the case of device B and D (Supp Figs.~\ref{fig:supp2}c,e), we observe clearly the linear regions of the qubit dispersion and that allows us to fit the linear Stark shift terms well. In the case of the other devices, where we typically observe only one side of the qubit dispersion, we either fit only one of the Stark shift terms for simplicity (in the case of Supp Fig.~\ref{fig:supp2}a,b) or none at all (in the case of Supp Fig.~\ref{fig:supp2}d). In these cases, having the linear Stark shift term $\eta_\mathrm{A,B}$ in the model can introduce an additional degree of freedom which negatively impacts the fit result. 

Similarly, the amplitude of electric driving $\Omega_\mathrm{AC}$ is in general required as an additional fitting parameter, except in the case of device D, where we experimentally estimate the AC voltage amplitude based on the impact of a microwave pulse on current profile of the device SET (shown in extended data, Fig.~1).

The parameter $c$ is treated as a free parameter to account for any constant offsets in the PESOS maps due to state preparation and measurement (SPAM) errors. We also have $\Delta E_\mathrm{Z}$ accounting for differences in Zeeman energies, which is typically on the order of tens of megahertz. 

Finally, we have the spin-orbit coupling parameters, which are the spin-dependent and spin-flip coupling terms respectively ($\Delta_\mathrm{sd}$ and $\Delta_\mathrm{sf}$). While these parameters are in general different, they are expected to be on the same order of magnitude. For the ease of fitting, we have them to be equal in magnitude in our model. The detailed justification for this will be given in the next section.

The primary result from the fits are the energy diagrams. The four-level model provides a reasonable description of the experiments for parameters extending across a few orders of magnitude. Not all parameters are uniquely set by this model due to either large uncertainties or weak dependence of the experimental results on certain parameters. However, the extracted parameters do lie within the expected bounds from either previous independent experiments or first principles calculations in the literature. The extracted magnitude of the electric drive corresponds to the maximum Rabi frequency shown in Supp Figs.~\ref{fig:supp2}(f-j). The difference in qubit frequencies are on the order of tens of megahertz (Table~\ref{table:fittedvals}) as expected for our devices \cite{yoneda2021coherent,tanttu2019,ruskov2018,ferdous2018interface}. For device C, the larger difference in $g$ factors resulting in a $\Delta E_\mathrm{Z}$ one order of magnitude larger than the rest could be due to the two orbitals having different shapes, such that the surface roughness along the interface impacts them in significantly different ways. The Stark shifts $\eta_\mathrm{A,B}$ range from a few MHz/V to hundreds of MHz/V. The lower bound is typical for dots without orbital degeneracy \cite{yoneda2021coherent,tanttu2019}, while the higher Stark shifts can be understood in terms of the sudden transition between $\ket{\mathrm{A}}$ and $\ket{\mathrm{B}}$. Confirming the physical significance of the remaining parameters $\Delta$ and $\Delta_\mathrm{sd/sf}$ requires a model of the orbitals involved, which we develop in the next section.

\subsection{Microscopic Origin of $\Delta$}
\label{subsec:deltaorigin}

The parameter $\Delta$ in our model describes how the orbitals $\mathrm{A}$ and $\mathrm{B}$ are \textit{coupled} to each other. Examples could include the case where $\mathrm{A}$ and $\mathrm{B}$ represent the $s$ and $p$ orbitals in an approximately harmonic confinement potential, in which case $\Delta$ stems from any small anharmonicity. Another example would see $p_x$ and $p_y$ states coupled through quadrupolar deformation of the dot. Often the two orbitals will also have a different valley composition, which means that the precise estimation of $\Delta$ is limited by atomistic disorder that might affect the valley structure of the dot~\cite{hosseinkhani2020,tariq2019,boross2016, yang2013,gamble2016,culcer2010}. 

To explore the full capabilities of this formalism, it would require knowledge of the microscopic details of the quantum dot, making this calculation impractical. While it remains possible to infer information about the anti-crossing from $\Delta$, there is in fact a caveat here in this case, that is the interplay between $\Delta$ and the differential lever arm $\alpha_\mathrm{rel}$ means that to draw any meaningful conclusions, an independent measurement of $\alpha_\mathrm{rel}$ is required as already discussed.  

As such, we will only comment on device A in the (1,3) configuration because $\alpha_\mathrm{rel}$ has been measured independently here. We first note that the measured $\alpha_\mathrm{rel} = 1200~\mathrm{GHz/V} \approx 5/h~\mathrm{meV/V}$ as shown in Fig.~1d of the main text. suggests that the observed pair $\ket{\mathrm{A}},\ket{\mathrm{B}}$ are orbital states rather than valley states which have a $\alpha_\mathrm{rel}$ about an order of magnitude smaller ($\alpha_\mathrm{rel} \approx 0.6/h~\mathrm{meV/V}$ in ref. \onlinecite{yang2013}). This accords with the fact that pairs of valley states typically do not differ much in charge distribution \cite{hosseinkhani2021}, whereas it is obvious that different orbital states do. 

Now the extracted $\Delta = 12 \pm 1~\mathrm{GHz} \approx (50\pm 4)/h$ $\mu$eV, as shown in Table~\ref{table:fittedvals}, is quite small and suggests that the avoided crossing is set mostly by valley-orbit coupling. To see this, we first note that we usually observe valley splittings $E_\text{vs}$ in the order of hundreds of $\mu$eV \cite{yang2013}. Under the effective mass approximation (EMA) formalism, the valley-orbit splitting can be written in general as \cite{saraiva2011}
\begin{equation}
    E_\text{vo} = 2\left|\bra{A}e\phi(\mathbf{r})\ket{B}\right|
\end{equation}
Normally for $E_\text{vs}$ (inter-valley, intra-orbital), $\mathrm{A}$ and $\mathrm{B}$ have approximately the same envelopes leading to a large overlap. We now apply the same calculation but now taking $\mathrm{A}$ and $\mathrm{B}$ to have different envelopes as justified earlier. The lower overlap results in a lower valley-orbit coupling (inter-valley, inter-orbital). So it is not unreasonable that $\Delta$ is an order of magnitude smaller than $E_\text{vs}$. Thus, this analysis suggests that $\ket{\mathrm{A}},\ket{\mathrm{B}}$ are different orbital states (having different orbital shapes) that occupy different valleys, and the change in gate voltage allows for transitions between the orbital states. This explanation is natural for the case of dots occupied by three electrons if the shell and valley structure of the dot are well preserved across the different occupation numbers.

\subsection{Origin of $\Delta_\text{sd}$ and $\Delta_\text{sf}$}
\label{subsec:socorigin}

The coupling terms $\Delta_\mathrm{sd}$ and $\Delta_\mathrm{sf}$ arise from spin-orbit coupling. To see this, we start with the spin-orbit coupling Hamiltonian, which reflects the relatively low symmetry of the rough Si/SiO$_2$ interface \cite{ruskov2018,nestoklon2006,tanttu2019}
\begin{equation}
     H_\text{SO} = \alpha \underbrace{(\sigma_{\bar{x}}k_{\bar{y}}-\sigma_{\bar{y}}k_{\bar{x}})}_{h_\text{R}} + \beta \underbrace{(\sigma_{\bar{x}}k_x-\sigma_{\bar{y}}k_{\bar{y}})}_{h_\text{D}} 
\end{equation}
where the Rashba and Dresselhaus Hamiltonians ($h_\mathrm{R}$ and $h_\mathrm{D}$ respectively) are scaled by the corresponding coefficients $\alpha, \beta$. The effect of inter-valley spin-orbit coupling is integrated into these coefficients in our analysis. The operators here $\bm{\sigma}, \mathbf{k}$ are defined according to the underlying crystallographic axes, which are represented with bars over $x$, $y$ and $z$ to underscore that they refer to $[100]$, $[010]$ and $[001]$ directions. These directions are not necessarily aligned with either the dot symmetry axes or the external magnetic field, such that both terms may play a role on spin-dependent and spin-flip coupling. In our experiments, for instance, the dots ideally have an approximate mirror symmetry about the $[110]$ direction and an external magnetic field applied along the $[1\bar{1}0]$ direction, which would mean that Rashba and Dresselhaus effects participate equally in $\Delta_\mathrm{sd}$ and $\Delta_\mathrm{sf}$. In reality this scenario is further complicated by the presence of interface roughness, which in the case of amorphous thermal Si/SiO$_2$ $(001)$ interfaces has no particular crystallographic structure. This means that $A$ and $B$ orbitals may well have no special symmetry.

This means that only in fortuitous cases $\Delta_\text{sd}$ and $\Delta_\text{sf}$ would differ by a large factor. In most cases, the approximation $\Delta_\text{sf}=\Delta_\text{sd}$ is quantitatively acceptable. Qualitatively, this equality might introduce artificial symmetries to the Hamiltonian, such that an analysis of the impact of this approximation is warranted for any predictive analysis.

\subsection{Estimating $\Delta_\text{sd}$ from spin-orbit coupling}
\label{subsec:socestimate}

We now turn to estimating the magnitude of $\Delta_\text{sd}$. To find the matrix element $\bra{\mathrm{A}}k_i\ket{\mathrm{B}}$, we first consider the case where $\ket{\mathrm{A}},\ket{\mathrm{B}}$ are different orbital states, i.e. $s$ and $p$ (we will consider a mixture of orbital states later). Take $p$ along the most elongated axis in an elliptical dot potential (which is the most common case). Assuming harmonic oscillator states gives an analytical expression
\begin{equation}
    \langle s | k | p \rangle = -i\sqrt{\frac{m^*\omega}{2\hbar}}.
\end{equation}
Taking a typical orbital splitting of $\hbar\omega = 3~\mathrm{meV}$~\cite{yang2012,leon2020}, we obtain $\langle s | k | p \rangle = -0.06i \; \text{nm$^{-1}$}$. 

In general, $\ket{\mathrm{A}},\ket{\mathrm{B}}$ could have mixed orbital character due to the anharmonicity of the dot or interface roughness. So, denoting $r$ as the proportion of orbital mixing, we could have $\ket{A} = \sqrt{1-r}\ket{s} + \sqrt{r}e^{i\varphi}\ket{p}$ and $\ket{B} = -\sqrt{r}e^{-i\varphi}\ket{s}+\sqrt{1-r}\ket{p}$ with the coefficients chosen to preserve orthonormality. Then the matrix element reads
\begin{equation}
    \bra{A}k\ket{B} = (1-r)\bra{s}k\ket{p} - re^{-2i\varphi}\bra{p}k\ket{s} = \bra{s}k\ket{p}\left[(1-r)-r e^{-i(2\varphi+\pi)}\right]
\end{equation}
By symmetry of the states, the absolute value $\overline{|\bra{\mathrm{A}}k\ket{\mathrm{B}}|}$ is symmetric under $r\rightarrow (1-r)$ and it is largest when $r=0$ or $1$, giving $\overline{|\bra{s}k\ket{p}|}$. The lower bound is given when $r=0.5$.
Averaging over the random phase $\varphi$, we find that,
\begin{equation}
    \overline{\left|\bra{s}k\ket{p}\right|} \geq \overline{\left|\bra{A}k\ket{B}\right|} \geq \overline{\left|\bra{s}k\ket{p}\right|} \; \overline{\left|0.5-0.5e^{-i(2\varphi+\pi)}\right|} = \frac{2}{\pi}\overline{\left|\bra{s}k\ket{p}\right|} \approx 0.64\overline{\left|\bra{s}k\ket{p}\right|}
\end{equation}
which shows that $\overline{\left|\bra{\mathrm{A}}k\ket{\mathrm{B}}\right|} $ does not depend strongly on the amount of orbital mixing $r$ and is always roughly the same order of magnitude.

A similar analysis may be performed for the case of $p_x$ and $p_y$ dots, in which case the direct transition induced by the vector $\{k_i\}$ is forbidden, but in second order it is allowable through the virtual coupling to $d$ orbitals.

We should also consider the case when $\ket{\mathrm{A}},\ket{\mathrm{B}}$ have different valley compositions, which is the most common case for very flat interfaces (energy levels ordered by valley splitting). Within the EMA formalism and only including the Bloch phase factor, the states appear approximately as $\psi_\text{A/B}(\mathbf{r}) = F_\mathrm{A/B}(\mathbf{r})e^{\pm ik_0z}$ where $F_\mathrm{A/B}$ is the envelope and the valleys occur at $\mathbf{k} = \pm (0,0,k_0)$, with $k_0=0.85 2\pi/a_0$. The matrix element becomes ($j=x,y$)
\begin{equation}
    \bra{A}k_j\ket{B} \approx -i\int d\mathbf{r} F^*_\mathrm{A}(\mathbf{r}) \frac{\partial F_\mathrm{B}(\mathbf{r})}{\partial r_j} e^{2ik_0z}
\end{equation}
We see that the valley phase likely attenuates the integral and we suppose that this attenuation could easily be a factor of $0.1 \sim 0.01$ or even less.

To compare theory with experiment, we note that in general the Dresselhaus contribution to the SOC is larger than the Rashba contribution \cite{tanttu2019,veldhorst2015,ruskov2018}. In ref. \onlinecite{tanttu2019} it was found that $\beta = 178(11) \times 10^{-13}$ eV cm while in ref. \onlinecite{ruskov2018} it was found that $\alpha - \beta$ is in the range from -300$\mu$eV nm to 800$\mu$eV nm. We thus estimate $|\alpha + \beta|  \approx (100 -- 1000)\mu$eV nm. If $\ket{\mathrm{A}},\ket{\mathrm{B}}$ belong to the same valley then we estimate $|\Delta_\text{sd}| \approx |\alpha + \beta| \times 0.1 \times\left|\langle s | k | p \rangle\right|= 10^{-1} \sim 10^0$ GHz as an order of magnitude estimate. For $\ket{\mathrm{A}},\ket{\mathrm{B}}$ inter-valley this becomes $\Delta_\text{sd} \approx 10^{-3} \sim 10^{-1}$ GHz. 

We see that the variation of the spin-orbit parameters as well as the valley-orbit character of the states leads to variation of $\Delta_\text{sd}$ among the devices. The estimates compare favorably with the fitted $\Delta_\text{sd}$ which are between $10^{-3} \sim 10^{-2}$ GHz (Table~\ref{table:fittedvals}). Our analysis suggests that the states $\ket{\mathrm{A}}, \ket{\mathrm{B}}$ are not states that have a valley index as a good quantum number and belong to the same valley. This is expected since typically in our dots the valley splitting is smaller than the orbital excitation energy.

For Device A in the (1,3) configuration, this conclusion is consistent with our previous remark - that the $\ket{\mathrm{A}},\ket{\mathrm{B}}$ here are in different orbital and valley states. In the other charge configuration (3,1), the smaller $\Delta_\text{sd}$ may be due to the different valley character that appears since the confinement potential and the interface that is probed by the different dot wavefunctions would be different here. For the maps in devices B and D, a similar conclusion holds because $\Delta_\text{sd}$ is of a similar magnitude. 

For device C, the value of $\Delta_\text{sd} = 0.06 \pm 0.08$ GHz $\approx 0.24\pm0.32~\mu\mathrm{eV}$ is the largest (although the confidence interval is not conclusive). This suggests we may be observing a mostly orbital anticrossing, with the valley composition of either orbital being compatible (valley interference not completely destructive). This is consistent with the large $\Delta \approx 1.2\pm0.8$ meV.

\begin{center}
\begin{table}[h!]
\resizebox{\textwidth}{!}{%
\begin{tabular}{ c c c c c c }
\hline\hline
\textbf{Parameters} & \textbf{Device A (1,3)} & \textbf{Device A (3,1)} & \textbf{Device B (3,1)} & \textbf{Device C (3,1)} & \textbf{Device D (3,1)} \\ 
\hline
$V_0$ (V) & -0.0528 & 0.36 & 1.56 & 0.91 & 1.6234 \\
$\alpha_\mathrm{rel}$ (THz/V) & 1.2 & 1.2 & 1.5 & 1.5 & 1.5 \\
$\Delta$ (GHz) & $12\pm1$ & $38\pm3$ & $85\pm3$ & $279\pm200$ & $1.3\pm0.3$ \\  
$\Delta_\mathrm{sd}$ or $\Delta_\mathrm{sf}$ (MHz) & $-27\pm5$ & $0\pm0.002$ & $-7.5\pm0.2$ & $-60\pm80$ & $-1.6\pm0.9$ \\
$dE_\mathrm{Z}$ (MHz) & $27\pm7$ & $-9\pm9$ & $10.3\pm0.3$ & $-110\pm70$ & $19\pm1$ \\
$\eta_\mathrm{A}$ (MHz/V) & $-238\pm30$ & Not Fitted & $-2.4\pm0.4$ & Not Fitted & $64\pm30$ \\
$\eta_\mathrm{B}$ (MHz/V) & $826\pm100$ & $5\pm4$ & $1.7\pm0.3$ & Not Fitted & $17\pm20$ \\
$c$ (MHz) & $21\pm10$ & $-58\pm10$ & $25.8\pm0.3$ & $-60\pm100$ & $73\pm1$ \\
$\Omega_\mathrm{AC}$ (MHz) & $9.5\pm0.6$ & $11.3\pm0.5$ & $1.81\pm0.04$ & $5.6\pm0.4$ & Not Fitted \\
\hline\hline
\end{tabular}}
\caption{Fitted parameters of the four-level model}
\label{table:fittedvals}
\end{table}
\end{center}

\bibliographystyle{naturemag}
\bibliography{EDSRsupp}

%% file: extData.tex
\makeatletter
\renewcommand{\figurename}{Extended Data Figure}
\makeatother

\renewcommand{\tablename}{Extended Data Table}
\makeatletter
\renewcommand*{\fnum@table}{{\normalfont\bfseries \tablename~\thetable}}
\makeatother

\captionsetup[table]{name={Extended Data Table},labelsep=period,justification=raggedright,font=small}
\captionsetup[figure]{name={\bf{Extended Data Fig.}},labelsep=line,justification=raggedright,font=small}
\setcounter{figure}{0}

\setcounter{table}{0}


%
%
%
\begin{table*}[ht]
\caption{\textbf{Devices details} \newline
Device A was fabricated on an isotopically enriched silicon-28 substrate (50 ppm residual $^{29}$Si) \cite{becker2010enrichment}.
Devices B, C \& D were fabricated on an epitaxially grown, isotopically purified $^{28}$Si epilayer with a residual $^{29}$Si concentration of 800 ppm \cite{itoh2014isotope}.} 
\label{table:devices} 
\centering 
\begin{tabular}{c c c c c c c} 
\hline\hline 
Device & \begin{tabular}{@{}c@{}} $^{29}$Si \\ concentration \end{tabular} & Gate Materials & Electron Occupancy & \begin{tabular}{@{}c@{}}$B_{0}$-Field \\ (T) \end{tabular}  & \begin{tabular}{@{}c@{}}Microwave Carrier \\ Frequency (GHz) \end{tabular} & Mode of Driving \\ [0.5ex] 
\hline 
A & 50 ppm & TiPd & (1,3) \& (3,1) & 0.825  & 22.3  & antenna-based \\ 
B & 800 ppm & TiPd & (3,1) & 0.55  & 15.4  & antenna-based \\
C & 800 ppm & Al & (1,3) & 0.825  & 22.3  & antenna-based \\
D & 800 ppm & Al & (3,1) & 0.1 - 1.0  & 2.8 - 28  & gate-based\\ [1ex] 
\hline \hline 
\end{tabular}

\end{table*}

\begin{figure*}[ht!]
    \centering
    \includegraphics[width = 1\textwidth, angle = 0]{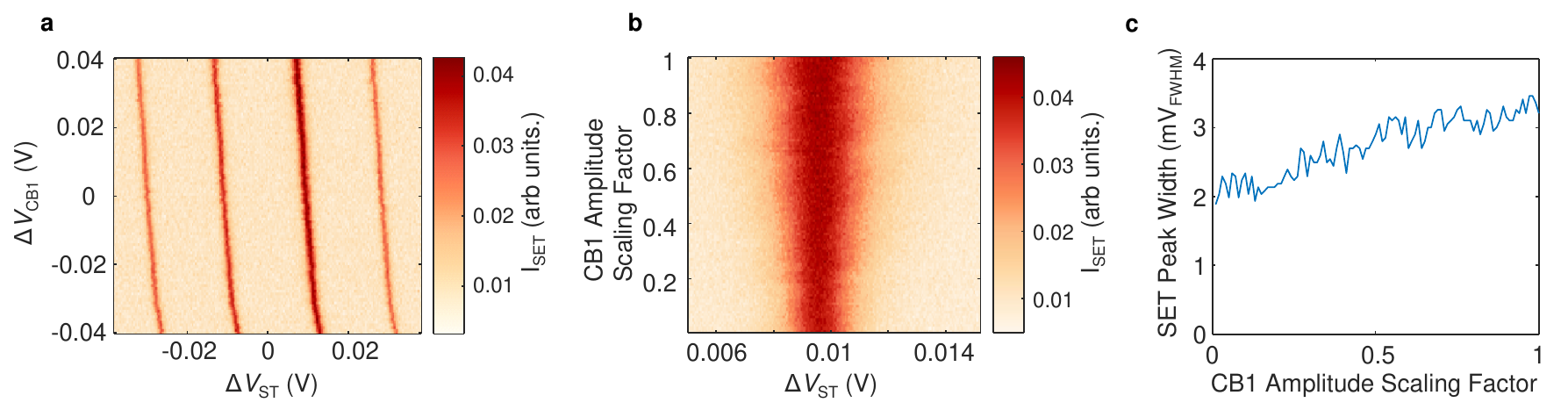}
    \caption{\textbf{Calibration of the microwave excitation amplitude}}
    The microwave signal applied to the CB1 gate of Device D is delivered through a transmission line that is poorly characterised for frequencies above 6 GHz. \textbf{a}, Here the line transmission is calibrated by first measuring the slope of Coulomb oscillations relative to the voltages on the sensor top gate and the CB1 gate,  which gives a relative lever arm of 14.8. Then, in \textbf{b} the width of a single Coulomb peak is measured against the microwave amplitude, scaled from a reference level of 0 dBm from the source. The fitted peak width trend in \textbf{c} gives a maximum amplitude at the CB1 gate of $50 \pm 2$~mV when multiplied by the relative lever arm.
    \label{fig:mw_amp}
\end{figure*}
 
\begin{figure*}[ht!]
    \centering
    \includegraphics[width = 0.85\textwidth, angle = 0]{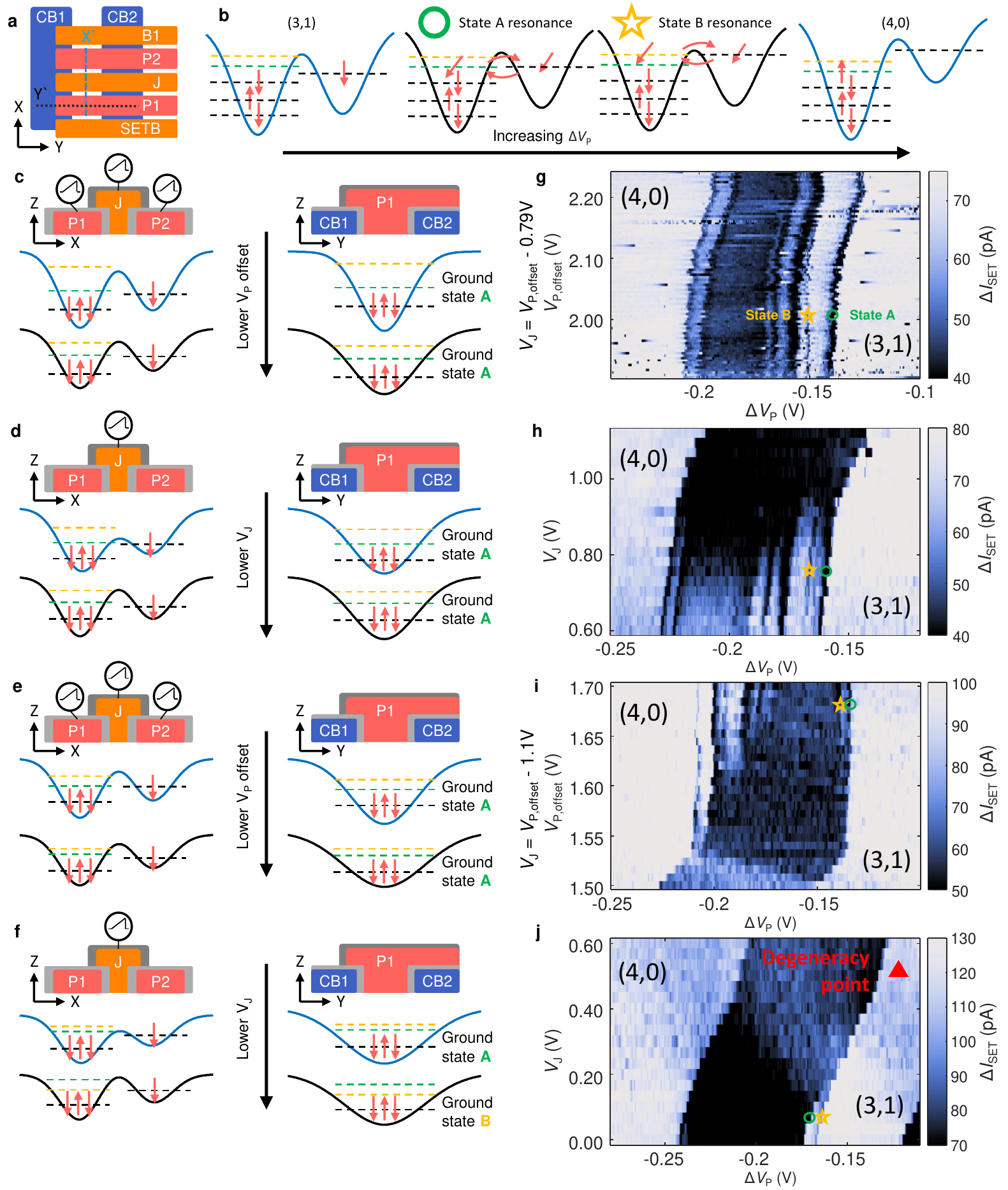}
    \caption{\textbf{Obtaining degeneracy on demand via excited state spectroscopy}}
    The protocol for tuning the energy spectrum of a new charge configuration to create a level degeneracy, resulting in a speed-up of the qubit Rabi frequency. The protocol utilises excited state spectroscopy, where a differential square wave excitation is applied to $\Delta V_{\mathrm P} = \Delta V_{\mathrm{P1}} = -\Delta V_{\mathrm{P2}}$, with in-phase charge movements detected using double lock-in down-conversion \cite{yang2011dynamically}. $\Delta V_{\mathrm P}$ is swept across an interdot charge transition to observe changes in the interdot tunnel rate associated with the population of excited states \cite{yang2012orbital}.
    \textbf{a}, Top-view schematic layout of the device gates. \textbf{b}, Schematic potential landscapes for increasing $\Delta V_{\mathrm P}$, progressing from deep in (3,1) to deep in (4,0), where in-between the ground state of the P2 dot crosses the level of multiple states for the 4th electron of the P1 dot. In \textbf{g} the two right-most features are designated as states A and B, representing the ground and 1st excited states of the P1 dot.
    To create a level degeneracy we observe the trend in the separation of states A and B whilst deforming the dots, and tune towards a point of zero separation. In \textbf{g}, $V_{\mathrm{P1}}$, $V_{\mathrm{P2}}$, and $V_{\mathrm{J}}$ are varied and denoted as $V_{\mathrm{P,offset}}$, with the expected potential deformation shown in \textbf{c}. The states A \& B appear to converge for decreasing $V_{\mathrm{P,offset}3}$, so we set $V_{\mathrm{P,offset}} = 1.7~V$ and in \textbf{d} \& \textbf{h} re-tune $V_{\mathrm{J}}$ to achieve desired tunnel rates.
    At the lower J value the excited states can be probed again. In \textbf{e} \& \textbf{i} we further lower $V_{\mathrm{P,offset}}$, and lose the visibility of the excited state due to high tunnel rates. In \textbf{f} \& \textbf{j} we again attempt to re-tune the interdot tunnel rate and again observe the states A \& B, here converging for higher $V_{\mathrm{J}}$, potentially a sign that states A \& B have already crossed. Now we perform a PESOS map near the red triangle where we anticipate a degeneracy, with results shown in \ref{fig:pesos}b in the main text. We note that the first PESOS map we took in the vicinity of the red triangle immediately showed a speed up of the qubit Rabi frequency. 
    \label{fig:obtain_degeneracy}
\end{figure*}

\begin{figure*}[ht!]
    \centering
    \includegraphics[width = 0.8\textwidth, angle = 0]{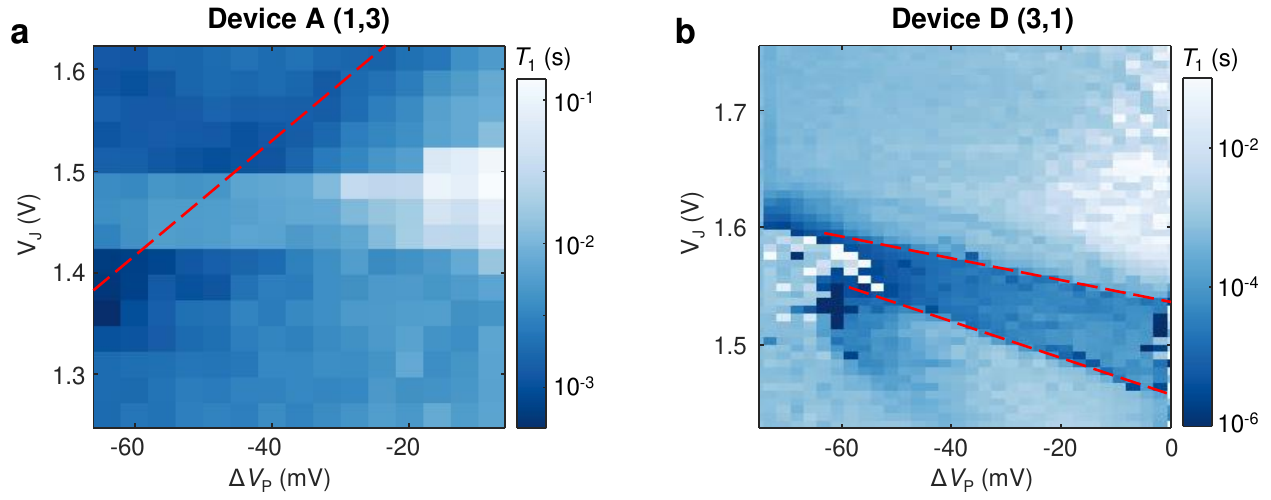}
    \caption{\textbf{Enhanced relaxation rate near level degeneracy}}
    The spin relaxation time $T_1$ measuring across the full target charge configurations for Device A (\textbf{a}) and Device D (\textbf{b}) in $V_{\rm J}$ and $\Delta V_{\rm P}$ gate space. The locations of level crossings are indicated as red dashed lines. This shows an enhanced spin relaxation rate near the point of level degeneracy in the respective systems. Interestingly, there are two visible transitions in Device D, between which the spin relaxation is faster than the background, potentially indicating the population of an alternative ground state.
    \label{fig:T1}
\end{figure*}

\begin{figure*}[ht!]
    \centering
    \includegraphics[width = 0.95\textwidth, angle = 0]{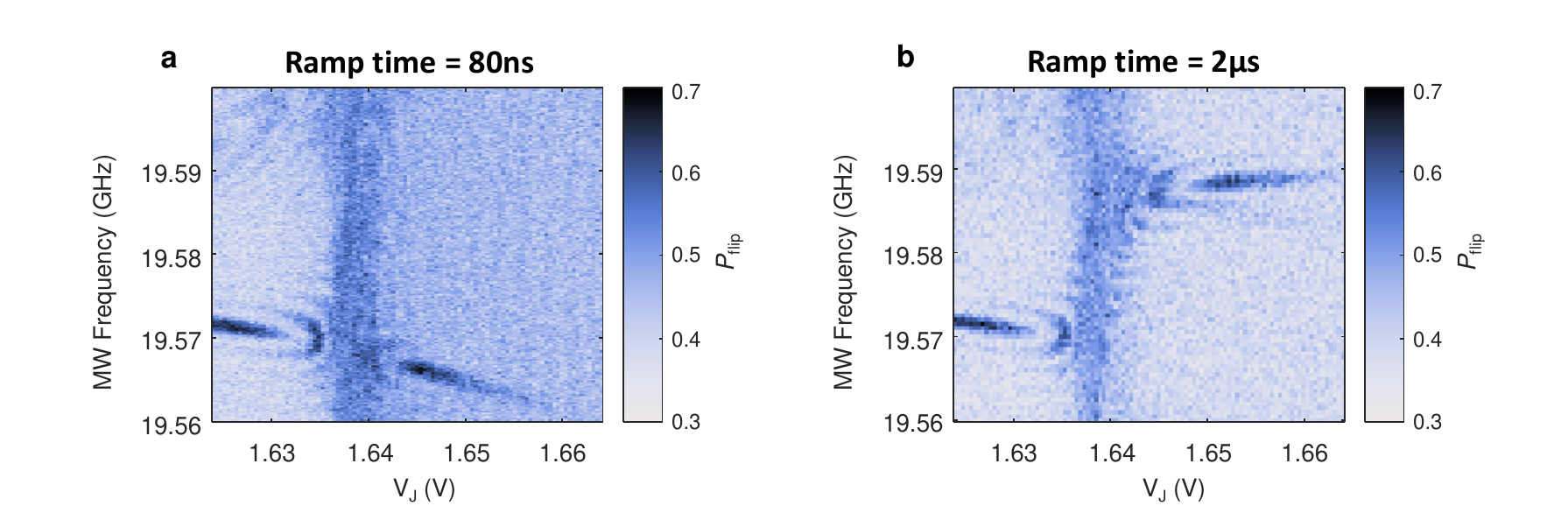}
    \caption{\textbf{Diabatic orbital excitation}}
    \textbf{a}, PESOS map in which the ramp time of the J-gate voltage is 80~ns. The resonance frequency continues on a linear trend, indicating that the orbital state is maintained when crossing the point of degeneracy at $V_{\rm J} = 1.638$~V.
    \textbf{b}, PESOS map in which the J-gate ramp time is increased to $2~{\mu}$s. The resonance frequency jumps after the degeneracy point, indicating population of a new orbital state via adiabatic passage across the level-crossing.
    \label{fig:diabatic}
\end{figure*}

\begin{figure*}[ht!]
    \centering
    \includegraphics[width = 0.8\textwidth, angle = 0]{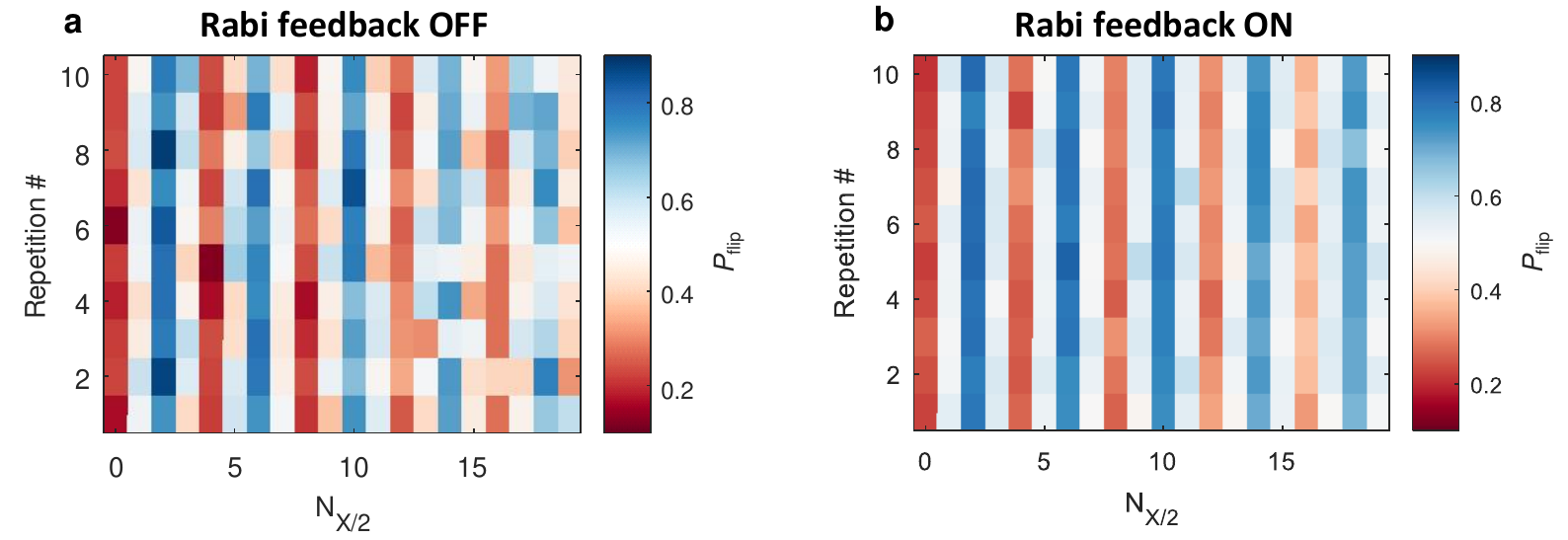}
    \caption{\textbf{Rabi Frequency Compensation}}
    \textbf{a} and \textbf{b} show Rabi oscillations discretised into ${\pi}/2$ rotations and repeated 10 times to observe instability, with Rabi frequency feedback on, and off, respectively. The feedback protocol measures $P_{\rm flip}$ for 5, 7, 9, and 11 ${\pi}/2$ pulses, each of which should result in a 0.5 flip proportion in the ideal case. It then applies a correction to the microwave amplitude of $\delta A_{\rm MW} = - gain * [P_5 - P_7 + P_9 - P_{11}]$, where $P_n$ is the measured spin flip proportion after $n$ ${\pi}/2$ rotations.
    \label{fig:rabifb}
\end{figure*}

\begin{figure*}[ht!]
    \centering
    \includegraphics[width = 0.8\textwidth, angle = 0]{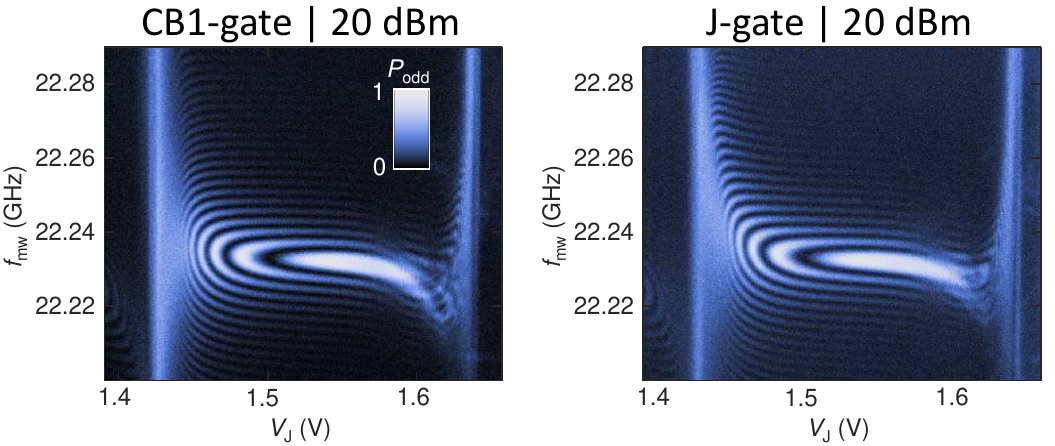}
    \caption{\textbf{Comparison of driving gates for EDSR}}
    \textbf{a}, PESOS map acquired with microwave drive applied to the CB1 gate.
    \textbf{b}, A PESOS map acquired using the same configuration as for \autoref{fig:qubit}, with microwave drive applied to the J-gate.
    Both measurements use the same source power of 20 dBm, however variation in line losses due to different electronics may result in differences in electric field amplitude applied to the CB1- and J-gates.
    \label{fig:drivegate}
\end{figure*}